\documentstyle[prd,preprint,tighten,aps,eqsecnum,floats,amssymb,amsfonts,newlfont,epsfig]{revtex} 
\begin{document} 
\draft 
\preprint{\vbox{ \hbox{hep-ph/0001101}   
\hbox{FTUV/00-02}  
\hbox{IFIC/00-02} 
\hbox{DFTT 69/99} }} 
\title{Four--Neutrino Oscillation Solutions of the Solar Neutrino Problem} 
\author{C. Giunti$^1$, M.\ C.\ Gonzalez-Garcia$^{2}$ and 
C. Pe\~na-Garay$^{2}$ } 
\address{\sl $^1$ INFN, Sez. di Torino, and Dip. di Fisica Teorica, 
Univ. di Torino, I--10125 Torino, Italy\\ 
$^2$  Instituto de F\'{\i}sica Corpuscular (IFIC) CSIC--Universidad de 
Valencia,\\
     Edificio Institutos de Paterna, Apartado 2085, 46071 Valencia}
\maketitle 
\begin{abstract} 
\baselineskip 0.5 cm 
We present an analysis of the neutrino oscillation  
solutions of the solar neutrino problem in the framework of  
four--neutrino  mixing where a sterile neutrino is added to the three  
standard ones. We perform a fit to the full data set 
corresponding to the 825-day Super--Kamiokande data sample as well as 
to Chlorine, GALLEX and SAGE and Kamiokande experiments. In our analysis  
we use all measured total event rates as well as all Super--Kamiokande data  
on the zenith angle dependence and the recoil electron energy spectrum. 
We consider both transitions via the  
Mikheyev--Smirnov--Wolfenstein (MSW) mechanism as well as oscillations 
in  vacuum (just-so) and  find the allowed solutions for different 
values of the additional mixing angles.  
This framework permits transitions into active or sterile neutrinos 
controlled by the additional parameter 
$\cos^2(\vartheta_{23}) \cos^2(\vartheta_{24})$   
and contains as limiting cases the pure  $\nu_e$--active and  
$\nu_e$--sterile neutrino oscillations.  
We discuss the maximum allowed values of this additional mixing  
parameter for the different solutions. 
As a particularity, we also show that for MSW transitions there 
are solutions at 99 \% CL at $\vartheta_{12}$ mixing angles greater than  
$\frac{\pi}{4}$ and that the best fit point for the zenith angle distribution  
is in the second octant. 
\end{abstract} 
\newpage 
 
\section{Introduction} 
 
Solar neutrinos were first detected already three decades ago  
in the Homestake experiment \cite{homestake0}  
and from the very beginning it was pointed out the puzzling issue of  
the deficit in the observed rate as compared to the theoretical  
expectation based on the standard solar model \cite{SSMold}  
with the implicit assumption that neutrinos created in 
the solar interior reach the Earth unchanged, i.e. they are massless 
and have only standard properties and interactions.  
This discrepancy led to a change in the 
original goal of using solar neutrinos to probe the properties of the 
solar interior towards the study of the properties of the neutrino 
itself and it triggered an intense activity both theoretical as well 
as experimental, with new measurements being proposed in order to 
address the origin of the deficit.  
 
On the theoretical side, enormous progress has been done in the  
improvement of solar modeling and 
calculation of nuclear cross sections. For example,  
helioseismological observations 
have now established that diffusion is occurring and by now most solar 
models incorporate the effects of helium and heavy element 
diffusion~\cite{Bahcall:1997qw,Bahcall:1995bt}. 
>From the experimental point of view the situation is now much 
richer. Four additional experiments to the original Chlorine experiment at  
Homestake \cite{homestake} have also detected solar neutrinos: 
the radiochemical Gallium experiments on $pp$ neutrinos, GALLEX 
\cite{gallex} and SAGE~\cite{sage}, and the water Cerenkov  
detectors Kamiokande~\cite{kamioka} and Super--Kamiokande 
\cite{sk1,sk99}. The latter have been able not only to confirm the original  
detection of solar neutrinos at lower rates than predicted by standard 
solar models, but also to demonstrate directly that the neutrinos come 
from the sun by showing that recoil electrons are scattered in the 
direction along the sun-earth axis. Moreover, they have also provided 
us with good information on the time dependence of the event rates during  
the day and night, as well as a measurement of the recoil electron energy  
spectrum. After 825 days of operation, Super--Kamiokande has also presented 
preliminary results on the seasonal variation of the neutrino event 
rates, an issue which will become important in discriminating the MSW 
scenario from the possibility of neutrino oscillations in 
vacuum~\cite{ourseasonal,v99}. At the present stage, the quality 
of the experiments themselves and the  
robustness of the theory give us confidence that in order to  
describe the data one must depart from the Standard Model (SM) of particle  
physics interactions by  
endowing neutrinos with new properties. In theories beyond the SM,  
neutrinos may naturally have new properties, 
the most generic of which is the existence of mass.   
It is undeniable that the most popular explanation of the solar  
neutrino anomaly is in terms of neutrino masses and mixing leading to neutrino 
oscillations either in {\sl vacuum}~\cite{Glashow:1987jj} or via the 
matter-enhanced {\sl MSW mechanism}~\cite{msw}. 
 
On the other hand, together with the results from the solar neutrino  
experiments we have two more evidences pointing out towards the existence of  
neutrino masses and mixing: the atmospheric neutrino data and the LSND 
results. The first one can be summarize in the existence of a long-standing  
anomaly between the predicted and observed $\nu_\mu/\nu_e$ ratio of the  
atmospheric neutrino fluxes \cite{atmexp}. In this respect it has been of  
crucial relevance the confirmation by the Super--Kamiokande collaboration  
\cite{sk99,atmSK}  
of the atmospheric neutrino zenith-angle-dependent deficit, which strongly  
indicates towards the existence of $\nu_\mu$ conversion.  
In addition to the solar and atmospheric neutrino results from underground 
experiments, there is also the indication for 
neutrino oscillations in the $\bar\nu_\mu \rightarrow \bar\nu_e$ channel  
by the LSND experiment~\cite{lsnd}. 
All these experimental results can be accommodated 
in a single neutrino oscillation framework only if there are at least 
three different scales of neutrino mass-squared differences. 
The simplest case of three independent 
mass-squared differences 
requires the existence of a light sterile neutrino, 
{\it i.e.} one whose interaction with 
standard model particles is much weaker 
than the SM weak interaction, 
so it does not affect the invisible Z decay  
width, precisely measured at LEP   
\cite{four-models,four-phenomenology,BGG-AB,Barger-variations-98,DGKK-99}. 
 
In this paper we present an analysis of the neutrino oscillation  
solutions of the solar neutrino problem in the framework of  
four--neutrino  mixing where a sterile neutrino is added to the three  
standard ones. We perform a fit of the full data set 
corresponding to the 825-day Super--Kamiokande data sample as well as 
the data of the Chlorine, GALLEX and SAGE experiments. 
In our analysis we use all 
measured total event rates and all Super--Kamiokande data on 
the zenith angle dependence and the recoil electron energy spectrum. 
We consider both transitions via the  
Mikheyev--Smirnov--Wolfenstein (MSW) mechanism as well as oscillations 
in  vacuum (just-so) and  find the allowed solutions for different 
values of the additional mixing angles. Our analysis contains as limiting  
cases the pure  $\nu_e$--active and $\nu_e$--sterile neutrino oscillations.  
We discuss the maximum allowed values of the additional mixing  
angles for which the different solutions are allowed. 
 
The outline of the paper is as follows. In Sec.~\ref{formalism} 
we summarize the main expressions for the neutrino oscillation formulas 
that we use in the analysis of solar neutrino data which take into account 
matter effects in the case of the MSW solution of the solar neutrino problem. 
We also present some improvement 
concerning the calculation of the regeneration 
of solar $\nu_e$'s in the Earth in Appendix~\ref{append}.  
Section ~\ref{sec:data} contains the summary of our calculations for 
the predictions of the different observables. Our quantitative results 
for the analysis of the four--neutrino oscillation parameters  
are given in Sec.~\ref{sec:fits}. Finally in Sec.~\ref{conclu} 
we summarize and discuss briefly our conclusions. 
 
\section{Four--Neutrino Oscillations} 
\label{formalism} 
 
In this paper we consider the two four-neutrino schemes 
that can accommodate the results of all neutrino oscillation experiments 
\cite{BGG-AB,Barger-variations-98}: 
\begin{equation} 
\mbox{(A)} 
\qquad 
\underbrace{ 
(\overbrace{m_1 < m_2}^{\Delta{m}^2_{\text{sun}}}) 
> 
(\overbrace{m_3 < m_4}^{\Delta{m}^2_{\text{atm}}}) 
}_{\Delta{m}^2_{\text{SBL}}} 
\,, 
\qquad 
\mbox{(B)} 
\qquad 
\underbrace{ 
(\overbrace{m_1 < m_2}^{\Delta{m}^2_{\text{sun}}}) 
< 
(\overbrace{m_3 < m_4}^{\Delta{m}^2_{\text{atm}}}) 
}_{\Delta{m}^2_{\text{SBL}}} 
\,. 
\label{AB} 
\end{equation} 
In both these mass spectra 
there are two pairs 
of close masses separated by a gap of about 1 eV 
which gives the mass-squared difference 
$ \Delta{m}^2_{\text{SBL}} = \Delta{m}^2_{41} $ 
responsible for the 
short-baseline (SBL) oscillations observed in the LSND experiment 
(we use the common notation 
$\Delta{m}^2_{kj} \equiv m_k^2 - m_j^2$). 
We have ordered the masses in such a way that 
in both schemes 
$ \Delta{m}^2_{\text{sun}} = \Delta{m}^2_{21} $ 
produces solar neutrino oscillations 
and 
$ \Delta{m}^2_{\text{atm}} = \Delta{m}^2_{43} $ 
is responsible for atmospheric neutrino oscillations. 
With this convention, 
the data of solar neutrino experiments can be analyzed using the 
neutrino oscillation formalism presented in Ref.~\cite{DGKK-99}, 
that takes into account matter effects. 
In this section we present the neutrino oscillation formulas 
that we use in the analysis of solar neutrino data. 
The transition probabilities that take into account 
matter effects in the case of the MSW solution of the solar neutrino problem 
have been derived in Ref.~\cite{DGKK-99}. 
Here we present some improvement 
concerning the calculation of the regeneration 
of solar $\nu_e$'s in the Earth 
(see the Appendix~\ref{append}). 
 
In four-neutrino schemes 
the flavor neutrino fields 
$\nu_{\alpha L}$ 
($\alpha=e,s,\mu,\tau$) 
are related 
to the fields $\nu_{kL}$ of neutrinos with masses $m_k$ 
by the relation 
\begin{equation} 
\nu_{\alpha L} = \sum_{k=1}^4 U_{\alpha k} \, \nu_{kL} 
\qquad 
(\alpha=e,s,\mu,\tau) 
\,, 
\label{mixing} 
\end{equation} 
where $U$ is a $4{\times}4$ unitary mixing matrix, 
for which we choose the parameterization 
\begin{equation} 
U 
= 
U_{34} \, U_{24} \, U_{23} \, U_{14} \, U_{13} \, U_{12} 
\,, 
\label{U4} 
\end{equation} 
where 
\begin{equation} 
(U_{ij})_{ab} 
= 
\delta_{ab} 
+ 
\left( \cos{\vartheta_{ij}} - 1 \right) 
\left( \delta_{ia} \delta_{ib} + \delta_{ja} \delta_{jb} \right) 
+ 
\sin{\vartheta_{ij}} 
\left( \delta_{ia} \delta_{jb} - \delta_{ja} \delta_{ib} \right) 
\label{rotation} 
\end{equation} 
represents a rotation in the 
$i$-$j$ 
$2\times2$ 
sector 
by an angle $\vartheta_{ij}$. 
In the parameterization (\ref{U4}) 
we have neglected, for simplicity, 
the possible presence of CP-violating phases. 
 
Since the negative results of the 
Bugey $\bar\nu_e$ disappearance experiment 
\cite{bugey} 
imply that 
$|U_{e3}|^2 + |U_{e4}|^2 \lesssim 3 \times 10^{-2}$ 
for $\Delta{m}^2_{\text{SBL}}$ 
in the LSND-allowed region 
$ 
0.2 \, \text{eV}^2 
\lesssim 
\Delta{m}^2_{\text{SBL}} 
\lesssim 
2 \, \text{eV}^2 
$, 
in the study of solar neutrino oscillations 
the matrices 
$U_{13}$ and $U_{14}$ 
can be approximated with the unit matrix 
({\it i.e.} $\vartheta_{13}=\vartheta_{14}=0$) 
and we obtain 
\begin{equation} 
U = U_{34} \, U_{24} \, U_{23} \, U_{12} 
\,. 
\label{U-sun} 
\end{equation} 

Explicitly, 
we have 
\begin{equation} 
U 
= 
\left( 
\begin{array}{cccc} \scriptstyle 
c_{12} 
& \scriptstyle 
s_{12} 
& \scriptstyle 
0 
& \scriptstyle 
0 
\\ \scriptstyle 
- s_{12} c_{23} c_{24} 
& \scriptstyle 
c_{12} c_{23} c_{24} 
& \scriptstyle 
s_{23} c_{24} 
& \scriptstyle 
s_{24} 
\\ \scriptstyle 
s_{12} 
( c_{23} s_{24} s_{34} 
+ s_{23} c_{34} ) 
& \scriptstyle 
- c_{12} 
( s_{23} c_{34} 
+ c_{23} s_{24} s_{34} ) 
& \scriptstyle 
c_{23} c_{34} 
- s_{23} s_{24} s_{34} 
& \scriptstyle 
c_{24} s_{34} 
\\ \scriptstyle 
s_{12} 
( c_{23} s_{24} c_{34} 
- s_{23} s_{34} ) 
& \scriptstyle 
c_{12} 
( s_{23} s_{34} 
- c_{23} s_{24} c_{34} ) 
& \scriptstyle 
- 
( c_{23} s_{34} 
+ s_{23} s_{24} c_{34} ) 
& \scriptstyle 
c_{24} c_{34} 
\end{array} 
\right) 
\,, 
\label{U} 
\end{equation} 
where 
$\vartheta_{12}$, 
$\vartheta_{23}$, 
$\vartheta_{24}$, 
$\vartheta_{34}$ 
are four mixing angles 
and 
$ c_{ij} \equiv \cos\vartheta_{ij} $ 
and 
$ s_{ij} \equiv \sin\vartheta_{ij} $. 
 
Since solar neutrino oscillations 
are generated by the mass-square difference 
between $\nu_2$ and $\nu_1$, 
it is clear from Eq.~(\ref{U}) 
that the survival of solar $\nu_e$'s 
mainly depends on the mixing angle 
$\vartheta_{12}$, 
whereas 
the mixing angles 
$\vartheta_{23}$ and $\vartheta_{24}$ 
determine the relative amount of transitions into sterile $\nu_s$ 
or 
active $\nu_\mu$ and $\nu_\tau$. 
Let us remind the reader that 
$\nu_\mu$ and $\nu_\tau$ 
cannot be distinguished in solar neutrino experiments, 
because their matter potential 
and their interaction in the detectors are equal, 
due only to neutral-current weak interactions. 
The active/sterile ratio 
and solar neutrino oscillations in general 
do not depend on 
the mixing angle 
$\vartheta_{34}$, 
that contribute only to the different mixings of 
$\nu_\mu$ and $\nu_\tau$, 
and depends on 
the mixing angles 
$\vartheta_{23}$ 
$\vartheta_{24}$ 
only through the combination 
$\cos{\vartheta_{23}} \cos{\vartheta_{24}}$. 
Indeed, 
from Eq.~(\ref{U}) 
one can see that the mixing of 
$\nu_s$ with $\nu_1$ and $\nu_2$ 
depends only on 
$\vartheta_{12}$ 
and the product 
$\cos{\vartheta_{23}} \cos{\vartheta_{24}}$. 
Moreover, 
instead of 
$\nu_\mu$ and $\nu_\tau$, 
one can consider 
the linear combinations 
\begin{equation} 
\left( 
\begin{array}{l} 
\nu_a 
\\ 
\nu_b 
\end{array} 
\right) 
= 
\left( 
\begin{array}{cc} 
- \sin{\vartheta} & - \cos{\vartheta} 
\\ 
\cos{\vartheta} & - \sin{\vartheta} 
\end{array} 
\right) 
\left( 
\begin{array}{cc} 
\sin{\vartheta_{34}} & \cos{\vartheta_{34}} 
\\ 
\cos{\vartheta_{34}} & - \sin{\vartheta_{34}} 
\end{array} 
\right) 
\left( 
\begin{array}{l} 
\nu_\mu 
\\ 
\nu_\tau 
\end{array} 
\right) 
\,, 
\label{ab1} 
\end{equation} 
with 
\begin{equation} 
\tan\vartheta 
= 
\frac{\sin{\vartheta_{24}}}{\tan{\vartheta_{23}}} 
\,. 
\label{ab2} 
\end{equation} 

The mixing of 
$\nu_a$ and $\nu_b$ 
with 
$\nu_1$ and $\nu_2$ 
is given by 
\begin{equation} 
U_{a1} 
= 
- s_{12} 
\sqrt{1 - c_{23}^2 c_{24}^2} 
\,, 
\quad 
U_{a2} 
= 
c_{12} 
\sqrt{1 - c_{23}^2 c_{24}^2} 
\,, 
\quad 
U_{b1} = U_{b2} = 0 
\,. 
\label{ab3} 
\end{equation} 

Therefore, the oscillations of solar neutrinos depend only on 
$\vartheta_{12}$ and the product 
$\cos{\vartheta_{23}} \cos{\vartheta_{24}}$. 
If $\cos{\vartheta_{23}} \cos{\vartheta_{24}} \neq 1$, 
solar $\nu_e$'s can transform 
in the linear combination $\nu_a$ of active $\nu_\mu$ and $\nu_\tau$. 
We distinguish the following limiting cases: 
\begin{itemize} 
\item{
 If $\cos{\vartheta_{23}} \cos{\vartheta_{24}} = 0$ then 
$U_{s1}=U_{s2}=0$, 
$ 
U_{a1} 
= 
- \sin{\vartheta_{12}} 
$, 
$ 
U_{a2} 
= 
\cos{\vartheta_{12}} 
$, 
corresponding to the limit of 
pure two-generation 
$\nu_e\to\nu_a$ transitions.} 
\item{
 If $\cos{\vartheta_{23}} \cos{\vartheta_{24}} = 1$ then 
$ 
U_{s1} 
= 
- \sin{\vartheta_{12}} 
$, 
$ 
U_{s2} 
= 
\cos{\vartheta_{12}} 
$ 
and 
$U_{a1}=U_{a2}=0$ 
and 
we have the limit of 
pure two-generation 
$\nu_e\to\nu_s$ transitions.} 
\end{itemize} 
 
Since 
the mixing of $\nu_e$ with $\nu_1$ and $\nu_2$ 
is equal to the one in the case of two-generations 
(with the mixing angle $\vartheta_{12}$), 
the mixing of $\nu_s$ with $\nu_1$ and $\nu_2$ 
is equal to the one in the case of two-generations 
times $\cos{\vartheta_{23}} \cos{\vartheta_{24}}$ 
and 
the mixing of $\nu_a$ with $\nu_1$ and $\nu_2$ 
is equal to the one in the case of two-generations 
times 
$\sqrt{ 1 - \cos^2{\vartheta_{23}} \cos^2{\vartheta_{24}} }$, 
it is clear that 
in the general case of simultaneous 
$\nu_e\to\nu_s$ 
and 
$\nu_e\to\nu_a$ 
oscillations 
the corresponding probabilities are given by 
\begin{eqnarray} 
&& 
P^{\text{Sun}}_{\nu_e\to\nu_s} 
= 
\cos^2{\vartheta_{23}} \cos^2{\vartheta_{24}} 
\left( 1 - P^{\text{Sun}}_{\nu_e\to\nu_e} \right) 
\,, 
\label{Pes} 
\label{Pea} 
\\ 
&& 
P^{\text{Sun}}_{\nu_e\to\nu_a} 
= 
\left( 1 - \cos^2{\vartheta_{23}} \cos^2{\vartheta_{24}} \right) 
\left( 1 - P^{\text{Sun}}_{\nu_e\to\nu_e} \right) 
\,. 
\end{eqnarray} 

These expressions satisfy the relation of 
probability conservation 
$ 
P^{\text{Sun}}_{\nu_e\to\nu_e} 
+ 
P^{\text{Sun}}_{\nu_e\to\nu_s} 
+ 
P^{\text{Sun}}_{\nu_e\to\nu_a} 
= 
1 
$. 
 
If $\Delta{m}^2_{21}$ is in the MSW region 
($ 
10^{-8} \, \text{eV}^2 
\lesssim 
\Delta{m}^2_{21} 
\lesssim 
3 \times 10^{-4} \, \text{eV}^2 
$), 
the survival probabilities of 
solar $\nu_e$'s 
is given by 
\cite{DGKK-99} 
\begin{equation} 
P^{\text{Sun}}_{\nu_e\to\nu_e} 
= 
\frac{1}{2}+ 
\left( \frac{1}{2} - P_c \right) \cos{2\vartheta_{12}} \, \cos{2\vartheta_{12}^M} 
\,. 
\label{Pee} 
\end{equation} 

Here the angle $\vartheta_{12}^M$ 
is the effective mixing angle in matter corresponding 
to the vacuum mixing angle $\vartheta_{12}$ 
and given by 
\begin{equation} 
\tan 2\vartheta_{12}^M 
= 
\frac 
{ \tan 2\vartheta_{12} } 
{ 1 - A / \Delta{m}^2_{21} \cos{2\vartheta_{12}} } 
\,, 
\label{theta12m} 
\end{equation} 
with 
\begin{equation} 
A 
\equiv 
A_{CC} + \cos^2{\vartheta_{23}} \, \cos^2{\vartheta_{24}} \, A_{NC} 
\,. 
\label{A} 
\end{equation} 

The quantities $A_{CC}$ and $A_{NC}$ 
describe the matter effects 
and are given by 
\begin{equation} 
A_{CC} 
= 
2 \sqrt{2} G_F E N_e 
\,, 
\qquad 
A_{NC} 
= 
- \sqrt{2} G_F E N_n 
\,, 
\label{A-CC-NC} 
\end{equation} 
where 
$N_e$ and $N_n$ 
are, 
respectively, 
the number densities of electrons and neutrons in the medium, 
$E$ is the neutrino energy 
and 
$G_F$ is the Fermi constant. 
The effective mixing angle $\vartheta_{12}^M$ 
in Eqs.~(\ref{Pee}) and (\ref{Pes}) 
must be evaluated at the point of neutrino production inside of the Sun. 
The quantity 
$P_c$ 
in Eq.~(\ref{Pee}) 
is the crossing probability given by the usual two-generation formula 
(see \cite{BGG-review-98}) 
and the replacement of the two-generation expression for $A$ 
with that given in Eq.~(\ref{A}). 
 
During the night solar neutrinos cross the Earth before reaching the detector 
and 
regeneration of $\nu_e$'s is possible 
\cite{daynight}. 
In the four-neutrino schemes under consideration, 
the probabilities of $\nu_e\to\nu_e$ and $\nu_e\to\nu_s$ transitions 
after crossing the Earth are given by 
\cite{DGKK-99} 
\begin{equation} 
P^{\text{Sun+Earth}}_{\nu_e\to\nu_e} 
= 
P^{\text{Sun}}_{\nu_e\to\nu_e} 
+ 
\frac{ 
\left( 
1 - 2 P^{\text{Sun}}_{\nu_e\to\nu_e} 
\right) 
\left( P^{\text{Earth}}_{\nu_2\to\nu_e} - \sin^2{\vartheta_{12}} \right) 
} 
{ \cos{2\vartheta_{12}} } 
\,, 
\label{PSEee} 
\end{equation} 
\begin{equation} 
P^{\text{Sun+Earth}}_{\nu_e\to\nu_s} 
= 
P^{\text{Sun}}_{\nu_e\to\nu_s} 
+ 
\frac{ 
\left( 
2 P^{\text{Sun}}_{\nu_e\to\nu_s} 
- 
\cos^2{\vartheta_{23}} \cos^2{\vartheta_{24}} 
\right) 
\left( 
P^{\text{Earth}}_{\nu_2\to\nu_s} 
- 
\cos^2{\vartheta_{12}} \cos^2{\vartheta_{23}} \cos^2{\vartheta_{24}} 
\right) 
} 
{ \cos{2\vartheta_{12}} \cos^2{\vartheta_{23}} \cos^2{\vartheta_{24}} } 
\,. 
\label{PSEes} 
\end{equation} 
The 
probability of $\nu_e\to\nu_a$ transitions 
is given by the conservation of probability: 
$ 
P^{\text{Sun+Earth}}_{\nu_e\to\nu_a} 
= 
1 
- 
P^{\text{Sun+Earth}}_{\nu_e\to\nu_e} 
- 
P^{\text{Sun+Earth}}_{\nu_e\to\nu_s} 
$. 
 
The probabilities 
$P^{\text{Earth}}_{\nu_2\to\nu_e}$ 
and 
$P^{\text{Earth}}_{\nu_2\to\nu_s}$ 
in Eqs.~(\ref{PSEee}) and (\ref{PSEes}) 
can be calculated by integrating numerically the differential equation 
that describes the evolution of neutrino flavors in the Earth 
(see \cite{DGKK-99}) or by using the analytical solution assuming 
a step-function profile of the Earth matter density  
(see \cite{Akhmedov-parametric-99}) . 
However, we notice that the probabilities 
$P^{\text{Earth}}_{\nu_2\to\nu_e}$ 
and 
$P^{\text{Earth}}_{\nu_2\to\nu_s}$ 
are not independent, because, 
as shown in the Appendix~\ref{append}, 
they are related by 
\begin{equation} 
P^{\text{Earth}}_{\nu_2\to\nu_s} 
= 
\cos^2{\vartheta_{23}} \cos^2{\vartheta_{24}} 
\left( 1 - P^{\text{Earth}}_{\nu_2\to\nu_e} \right) 
\,. 
\label{relation} 
\end{equation} 
Therefore, 
in the analysis of solar neutrino data we need to calculate only 
$P^{\text{Earth}}_{\nu_2\to\nu_e}$. 
 
If $\Delta{m}^2_{21}$ 
is in the range of the vacuum oscillation solution 
of the solar neutrino problem 
($ 
10^{-11} \, \text{eV}^2 
\lesssim 
\Delta{m}^2_{21} 
\lesssim 
10^{-9} \, \text{eV}^2 
$), 
the survival probability of solar $\nu_e$ 
is given by the two-generation formula 
\begin{equation} 
P^{\text{Sun}}_{\nu_e\to\nu_e} 
= 
1 - \sin^2{2\vartheta_{12}} \sin^2 \frac{\Delta{m}^2_{21}L}{4E} 
\,, 
\label{PeeVO} 
\end{equation} 
where $E$ is the neutrino energy and $L$ 
is the Sun--Earth distance, 
whose seasonal variations must be taken into account.  
In this case there is no matter effect during neutrino propagation in the  
Earth. 
\section{Data and Techniques} 
\label{sec:data} 
 
In order to study the possible values of neutrino masses and mixing 
for the oscillation solution of the solar neutrino problem, we have used data 
on the total event rates measured in the Chlorine experiment at 
Homestake \cite{homestake}, in the two Gallium experiments GALLEX and 
SAGE \cite{gallex,sage} and in the water Cerenkov detectors Kamiokande and 
Super--Kamiokande shown in Table~\ref{rates}. Apart from the total event 
rates, we have in this last case 
the zenith angle distribution of the events and the electron recoil 
energy spectrum, all measured 
with their recent 825-day data sample~\cite{sk99}. Although, as discuss 
in Ref. \cite{lisi3} the inclusion of Kamiokande results does not affect 
the shape of the regions, because of the much larger precision of the  
Super--Kamiokande measurement, it is convenient to introduce it   
as in this way the number of degrees of freedom for the fit of 
the rates only is $4-3=1$ (instead of zero degrees of freedom), 
that allows the construction of a well--defined $\chi^2_{min}$ confidence  
level.  
 
For the  calculation of the theoretical expectations we use the BP98 standard  
solar model of Ref.~\cite{BP98}.  
The general expression of the expected event rate in the presence of 
oscillations in experiment $i$ in the four--neutrino  
framework is given by $R^{th}_i$ : 
\begin{eqnarray} 
R^{th}_i & = & \sum_{k=1,8} \phi_k 
\int\! dE_\nu\, \lambda_k (E_\nu) \times  
\Big[ \sigma_{e,i}(E_\nu)  \langle  
P_{\nu_e\to\nu_e}\rangle \label{ratesth} \\ 
& &                            + \sigma_{x,i}(E_\nu)  
\bigg(1-\langle P_{\nu_e\to\nu_e}\rangle  
-\langle P_{\nu_e\to\nu_s}\rangle \bigg)\Big] \nonumber. 
\end{eqnarray}   
where $E_\nu$ is the neutrino energy, $\phi_k$ is the total neutrino 
flux and $\lambda_k$ is the neutrino energy spectrum (normalized to 1) 
from the solar nuclear reaction $k$ with the 
normalization given in Ref.~\cite{BP98}. Here $\sigma_{e,i}$ 
($\sigma_{x,i}$) is the $\nu_e$ ($\nu_x$, $x=\mu,\,\tau$) interaction 
cross section in the Standard Model with the target 
corresponding to experiment $i$. 
For the Chlorine and Gallium experiments we use improved cross 
sections $\sigma_{\alpha,i}(E)$ $(\alpha = e,\,x)$ from 
Ref.~\cite{prod}. For the Kamiokande and Super--Kamiokande experiment  
we calculate the expected signal with the corrected cross section as  
explained below. $\langle P_{\nu_e\to\nu_\alpha} \rangle$ is  
the time--averaged $\nu_e$ survival probability. 
In case of MSW transitions $ P_{\nu_e\to\nu_e}$  
and $P_{\nu_e\to\nu_s}$ are given in Eqs.(\ref{PSEee}) and {(\ref{PSEes})} 
respectively.  
 
For vacuum oscillations we must 
include the effect of the Earth orbit eccentricity. The yearly averaged 
probability is obtained by averaging Eq.(\ref{PeeVO})
 with $L(t)=L_0 [ 1 - \varepsilon\cos 2\pi \frac{t}{T}]$:
\begin{equation} 
\langle P_{\nu_e\to\nu_e}\rangle= 
\langle P^{\text{Sun}}_{\nu_e\to\nu_e}\rangle 
=2 - \sin^2{2\vartheta_{12}} \left[ 1 - \cos\!\left(\frac{\Delta{m}^2_{21} 
L_0}{2E}\right) 
J_{0}\!\left( \frac{\varepsilon \Delta{m}^2_{21} L_0}{2E}\right)\right] 
\,, 
\label{PeeVOav} 
\end{equation} 
where $\varepsilon$ is the orbit eccentricity ($0.0167$), $L_0$ is the  
average Earth orbit radius ($1.496\times 10^8$ km) and $J_{0}(x)$ is the
Bessel function. 
We have also included in the fit the experimental results from the 
Super--Kamiokande Collaboration on the zenith angle distribution of 
events taken on 5 night periods and the day averaged value,  
which we graphically reduced from 
Ref.~\cite{sk99}. For MSW oscillations we compute the expected event rate  
in the period $a$ in the presence of oscillations as, 
\begin{eqnarray}  
R^{th}_{sk,a}  
& = & \frac{\displaystyle 1}{\displaystyle\Delta \tau_a} 
\int_{\tau(\cos\Phi_{min,a})}^{\tau(\cos\Phi_{max,a})}  d\tau 
\sum_{k=1,8} \phi_k\int\! dE_\nu\, \lambda_k (E_\nu) \times  
\Big[ \sigma_{e,sk}(E_\nu)  
\langle P_{\nu_e\to\nu_e} (\tau) \rangle   
\label{eq:daynight}\\  
& &+ \sigma_{x,sk}(E_\nu)  
\bigg(1-\langle P_{\nu_e\to\nu_e} (\tau) \rangle   
-\langle P_{\nu_e\to\nu_s} (\tau) \rangle \bigg)\Big] \nonumber 
\,, 
\end{eqnarray}   
where $\tau$ measures the yearly averaged length of the period $a$  
normalized to 1, so $\Delta\tau_a=\tau(\cos\Phi_{max,a})-\tau 
(\cos\Phi_{min,a})=$ 
.500, .086, .091, .113, .111, .099 for the day and five night periods. 
Notice that for vacuum oscillations there is no matter effect during neutrino  
propagation in the Earth. In this case 
$R^{th}_{sk,a}=R^{th}_{sk}$, as given in Eq.(\ref{ratesth}).   
The Super--Kamiokande Collaboration has also presented 
the results on the day-night 
variation in the form of a day-night asymmetry.  
Since the information included in the zenith angle dependence already 
contains the day-night asymmetry, we have not added the asymmetry as 
an independent observable in our fit. 
 
The Super-Kamiokande Collaboration has also measured the recoil 
electron energy spectrum.  In their published analysis \cite{sk1} 
after 504 days of operation they present their results for energies 
above 6.5 MeV using the Low Energy (LE) analysis in which the recoil 
energy spectrum is divided into 16 bins, 15 bins of 0.5 MeV energy 
width and the last bin containing all events with energy in the range 
14 MeV to 20 MeV.  Below 6.5 MeV the background of the LE analysis 
increases very fast as the energy decreases. Super--Kamiokande has 
designed a new Super Low Energy (SLE) analysis in order to reject this 
background more efficiently so as to be able to lower their threshold 
down to 5.5 MeV. In their 825-day data \cite{sk99} they have used the 
SLE method and they present results for two additional bins with 
energies between 5.5 MeV and 6.5 MeV. 
In our study we use the experimental results from the 
Super--Kamiokande Collaboration on the recoil electron spectrum divided in 
18 energy bins, including the results from the LE analysis for the 16 
bins above 6.5 MeV and the results from the SLE analysis for the two 
low energy bins below 6.5 MeV.   
The general expression of the expected rate in a bin in the presence of 
oscillations, $R^{th}$, is similar to that in Eq.(\ref{ratesth}), 
with the substitution of the cross sections with the corresponding 
differential cross sections 
folded with the finite energy resolution function of the detector 
and integrated over the electron recoil energy interval of the bin, 
$T_{\text {min}}\leq T\leq T_{\text {max}}$: 
\begin{equation} 
\sigma_{\alpha,sk}(E_\nu)=\int_{T_{\text {min}}}^{T_{\text {max}}}\!dT 
\int_0^{\frac{E_\nu}{1+m_e/2E_\nu}} 
\!dT'\,Res(T,\,T')\,\frac{d\sigma_{\alpha,sk}(E_\nu,\,T')}{dT'}\ . 
\label{sigma} 
\end{equation} 
The resolution function $Res(T,\,T')$ is of the form~\cite{sk1,Faid}: 
\begin{equation} 
Res(T,\,T') = \frac{1}{\sqrt{2\pi}(0.47 \sqrt{T'\text{(MeV)}})}\exp 
\left[-\frac{(T-T')^2}{0.44\,T' ({\text {MeV}})}\right]\ , 
\end{equation} 
and we take the differential cross section $d\sigma_\alpha(E_\nu,\,T')/dT'$  
from \cite{CrSe}. 
 
In the statistical treatment of all these data we perform a $\chi^2$ 
analysis for the different sets of data,  
following closely the analysis of Ref.~\cite{fogli-lisi} with the  
updated uncertainties given in Refs.~\cite{lisi3,BP98,prod}, as 
discussed in Ref.~\cite{oursolar}.  We thus define  
a $\chi^2$ function for the three set of observables  
$\chi^2_{\text {rates}}$, $\chi^2_{\text {zenith}}$, and  
$\chi^2_{\text {spectrum}}$ where in both $\chi^2_{\text {zenith}}$, and 
 $\chi^2_{\text {spectrum}}$ we allow for a free normalization in order to  
avoid double-counting with the data on the total event rate which is 
already included in $\chi^2_{\text {rates}}$.   
In the combinations of observables we define the 
$\chi^2$ of the combination as the sum of the different $\chi^2$'s.  
In principle such  analysis should be taken with a grain of salt as 
these pieces of information are not fully independent; in fact, they 
are just different projections of the double differential spectrum of 
events as a function of time and energy. Thus, in our combination we 
are neglecting possible correlations between the uncertainties in the 
energy and time dependence of the event rates. 
 
\section{Results} 
\label{sec:fits} 
As explained in Sec.~\ref{formalism}, for the mass scales 
invoked in the explanation of the atmospheric and LSND 
data and after imposing the strong constraints from  
the Bugey\cite{bugey} and CHOOZ \cite{chooz} reactor experiments, 
the relevant parameter space for solar neutrino oscillations 
in the framework of four--neutrino mixing is a three dimensional 
space in the variables $\Delta{m}^2_{21}$,  
$\vartheta_{12}$ and $\cos^2(\vartheta_{23})\cos^2(\vartheta_{24})\equiv  
c^2_{23} c^2_{24}$. 
As shown in Section~\ref{formalism}, 
the case $ c^2_{23} c^2_{24}=0$ corresponds to  
the usual two--neutrino oscillations $\nu_e\rightarrow \nu_a$ where 
$\nu_a$ is the admixture of $\nu_\mu$ and $\nu_\tau$ 
given in Eq.~(\ref{ab1}), thus  
an active neutrino. The other extreme case   
$ c^2_{23} c^2_{24}=1$ corresponds to the usual two--neutrino  
oscillations of $\nu_e$ into a pure sterile neutrino. 
 
In our choice of ordering the neutrino masses in the two schemes (\ref{AB}) 
the mass-squared difference $\Delta{m}^2_{21}$ 
is positive. The mixing angle $\vartheta_{12}$ 
can vary in the interval 
$0 \leq \vartheta_{12} \leq \frac{\pi}{2}$. 
In the case of vacuum oscillations, 
the transition probabilities are 
symmetric under the change 
$\vartheta_{12} \rightarrow \frac{\pi}{2}-\vartheta_{12}$ 
and each allowed value of $\sin^2(2\vartheta_{12})$ 
corresponds to two allowed values of 
$\vartheta_{12}$. 
On the other hand, 
in the case of the MSW solutions 
the transition probabilities are 
not invariant under the change 
$\vartheta_{12} \rightarrow \frac{\pi}{2}-\vartheta_{12}$ 
and resonant transitions are possible only for 
values of $\vartheta_{12}$ 
smaller than $\frac{\pi}{4}$. 
In the analysis of the observable rates, we present the results in the common 
plot of $\sin^2(2\vartheta_{12})$ due to the fact that the allowed region  
does not extend to $\vartheta_{12}>\frac{\pi}{4}$. But, when we include the  
rest of observables we remark that this is not the case, and we present the  
results 
as a function of $\sin^2(\vartheta_{12})$ (see \cite{Murayama}) showing that 
there is a portion of the space of parameters in the second octant  
of $\vartheta_{12}$ allowed at 99 \% CL. This enlarged parameter space is  
also used in Ref.~\cite{lisi3}. 
 
\begin{figure} 
\begin{center} 
\mbox{\epsfig{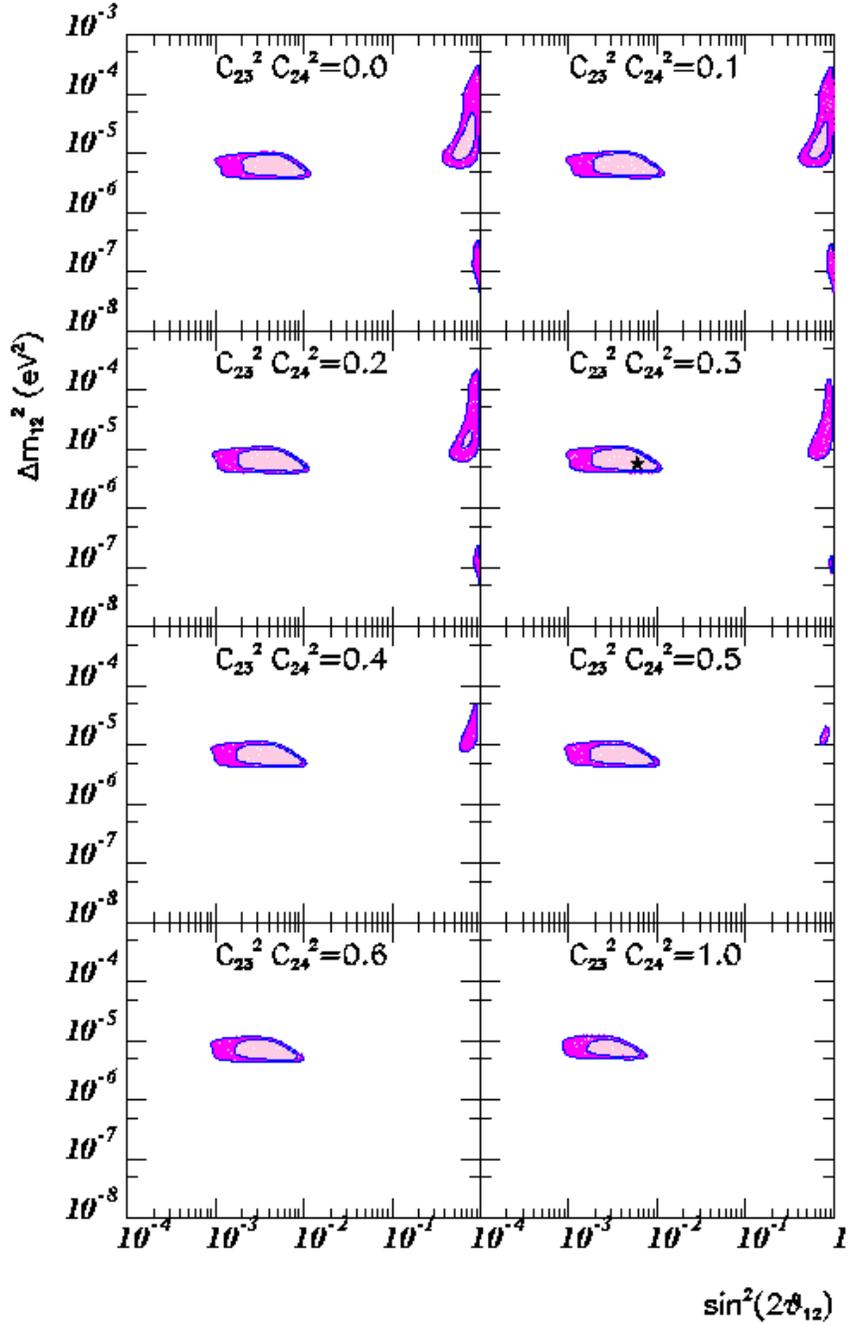}} 
\end{center} 
\caption{Allowed regions in  $\Delta{m}^2_{21}$ and $\sin^2(2\vartheta_{12})$  
for the MSW four--neutrino oscillations 
from the measurements of the total event rates at Chlorine, Gallium,  
Kamiokande and Super--Kamiokande (825-day data sample). The different panels 
represent the allowed regions at 99\% (darker) and 90\% CL (lighter)  
obtained as sections for fixed values of the mixing angles  
$c^2_{13} c^2_{23}$ of the three--dimensional volume 
defined by $\chi^2-\chi^2_{min}$=6.25 (90\%), 11.36 (99\%). 
The best--fit point in the three parameter space is  
plotted as a star.} 
\label{fig:msw_r} 
\end{figure}

\begin{figure} 
\begin{center} 
\mbox{\epsfig{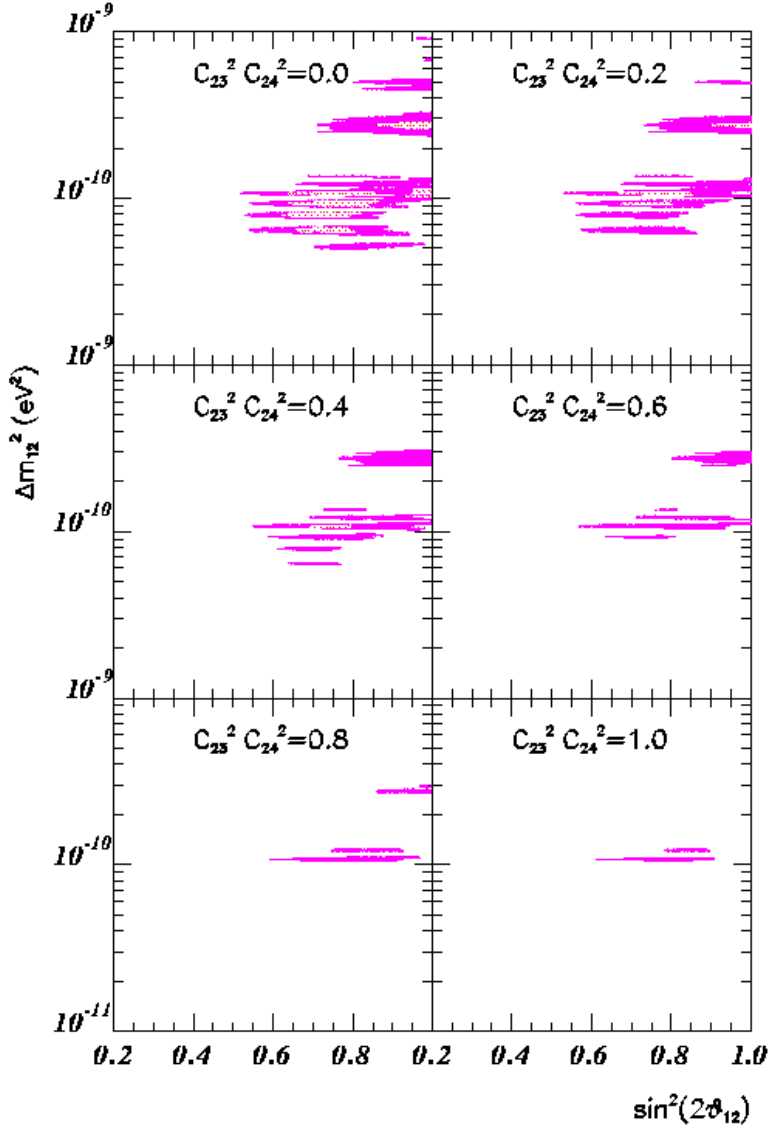}} 
\end{center} 
\caption{Allowed regions in  $\Delta{m}^2_{21}$ and $\sin^2(2\vartheta_{12})$  
for the vacuum four--neutrino oscillations 
from the measurements of the total event rates at Chlorine, Gallium,  
Kamiokande and Super--Kamiokande (825-day data sample). The different panels 
represent the allowed regions at 99\% (darker) and 90\% CL (lighter)  
obtained as sections for fixed values of the mixing angles  
$c^2_{13} c^2_{23}$ of the three--dimensional volume 
defined by $\chi^2-\chi^2_{min}$=6.25 (90\%), 11.36 (99\%) 
where $\chi^2_{min}$ is in the MSW region.} 
\label{fig:vac_r} 
\end{figure}

\begin{figure} 
\begin{center} 
\mbox{\epsfig{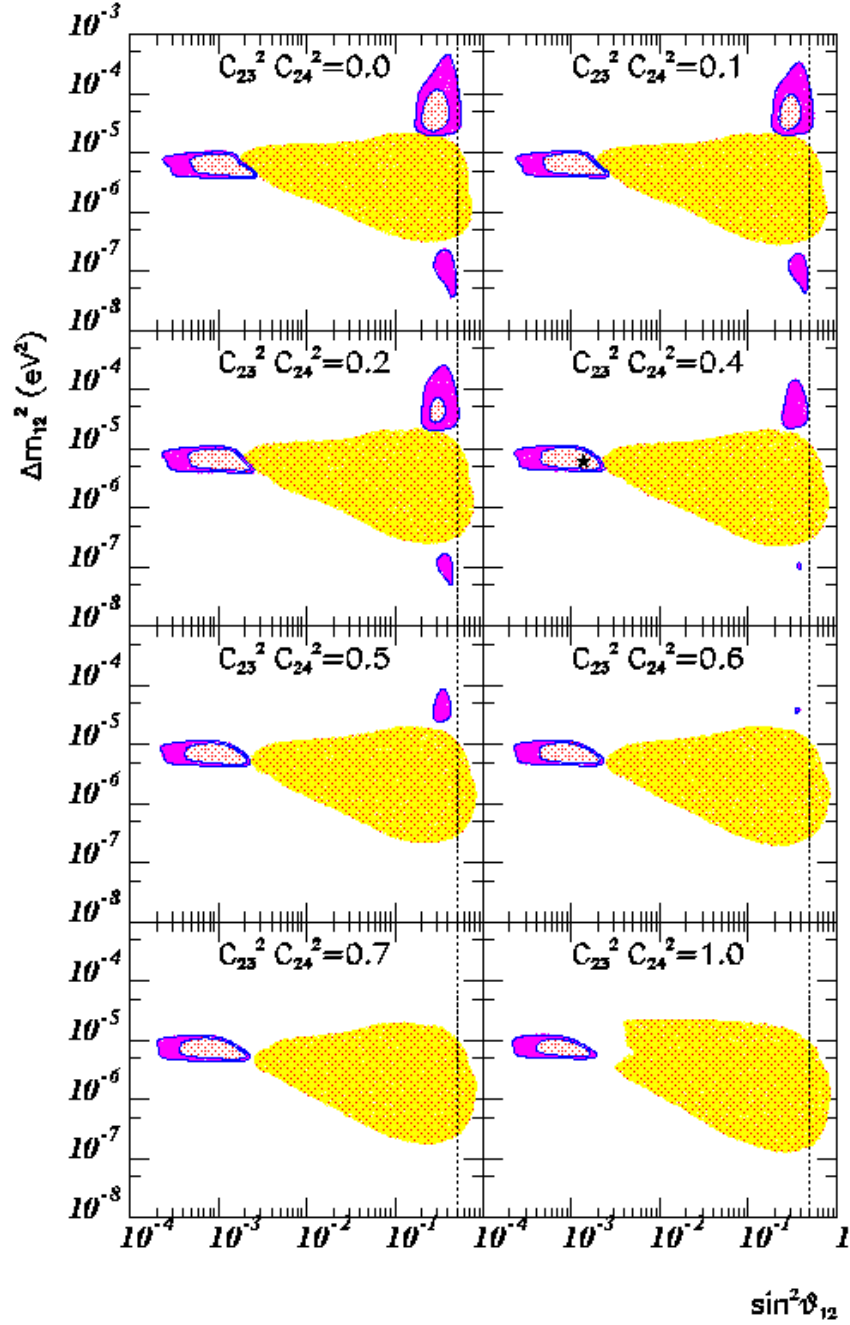}} 
\end{center} 
\caption{Allowed regions in  $\Delta{m}^2_{21}$ and $\sin^2 \vartheta_{12}$  
for the MSW four--neutrino oscillations 
from the measurements of the event rates and the Super--Kamiokande  
zenith angular dependence data. The shadowed area represents the  
excluded region at 99\% CL from the zenith angular data.}   
\label{fig:msw_rz} 
\end{figure}

\begin{figure} 
\begin{center} 
\mbox{\epsfig{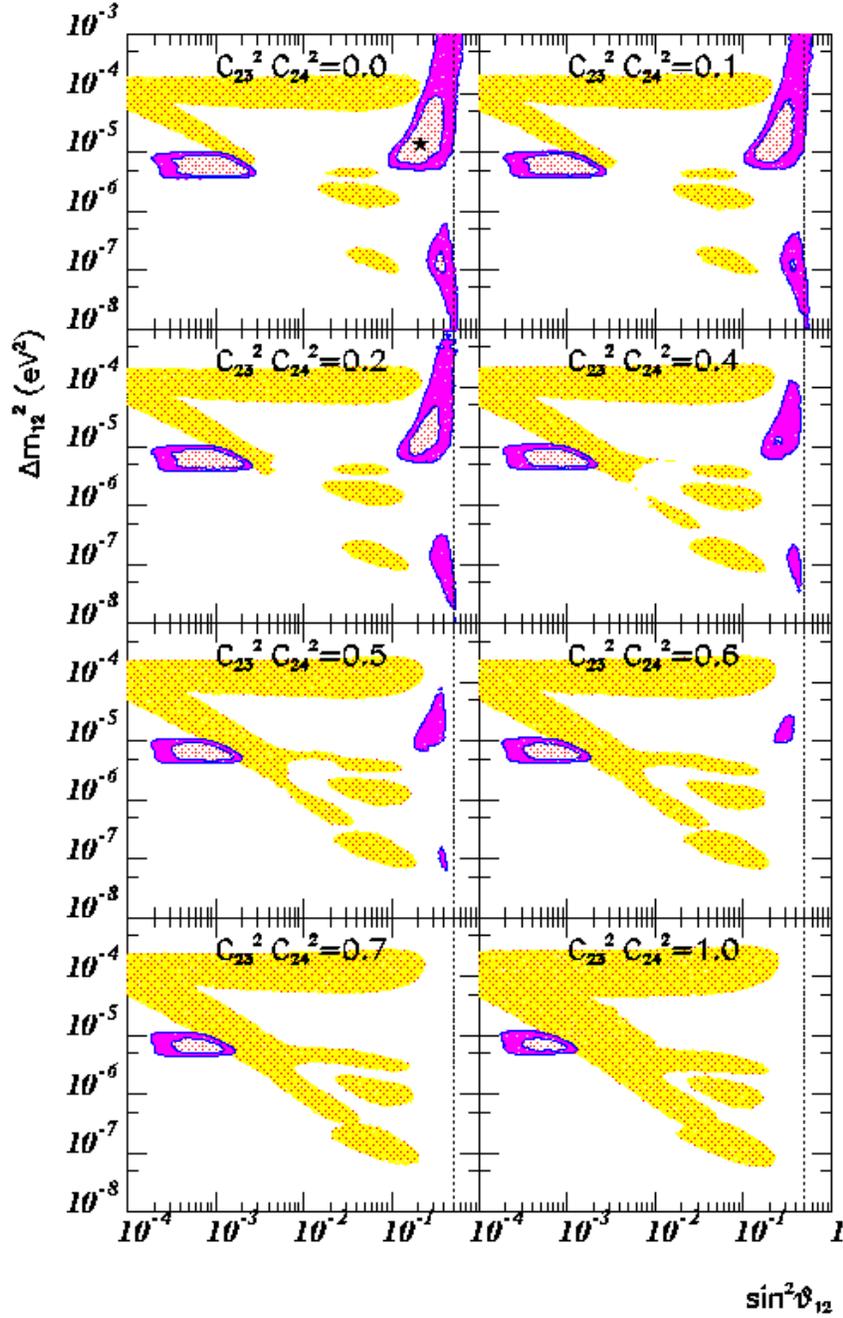}} 
\end{center} 
\caption{Same as Fig.{\protect\ref{fig:msw_rz}} but 
for the measurements of the event rates and the Super--Kamiokande  
recoil electron energy spectrum. 
The shadowed area represents the  
excluded region at 99\% CL from the energy spectrum.}   
\label{fig:msw_rs} 
\end{figure}

\begin{figure} 
\begin{center} 
\mbox{\epsfig{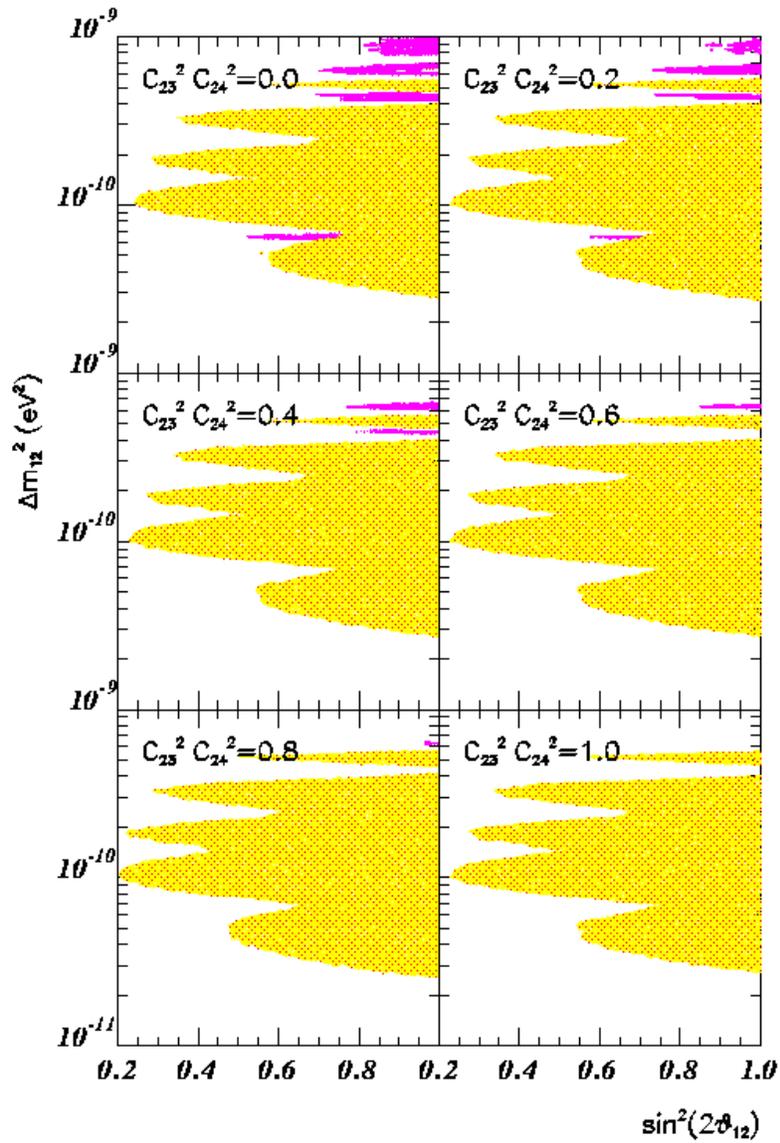}} 
\end{center} 
\caption{Same as Fig.{\protect\ref{fig:vac_r}} but  
for the measurement of the event rates and the Super--Kamiokande  
recoil electron energy spectrum. 
The shadowed area represents the  
excluded region at 99\% CL from the energy spectrum.}   
\label{fig:vac_rs} 
\end{figure}

\begin{figure} 
\begin{center} 
\mbox{\epsfig{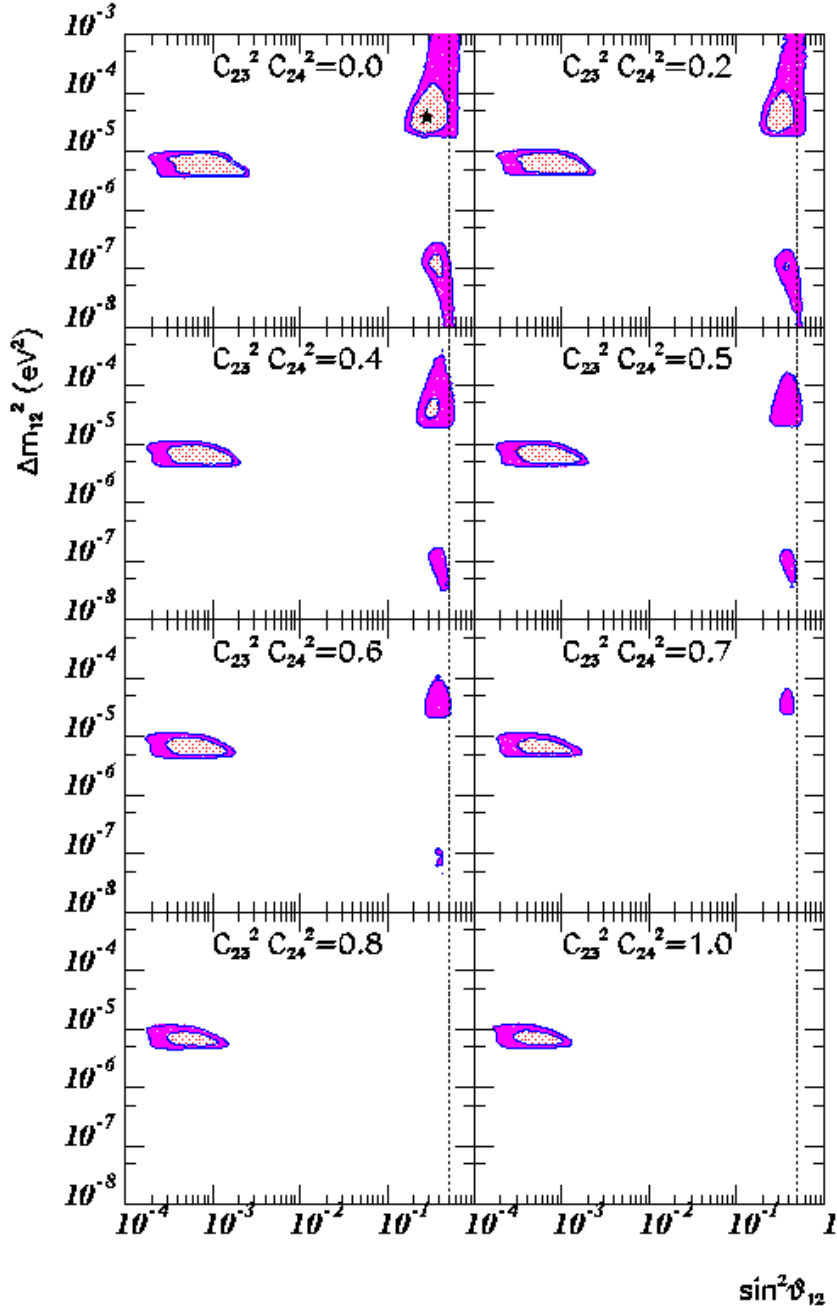}} 
\end{center} 
\caption{Results of the global analysis for the allowed regions in   
$\Delta{m}^2_{21}$ and $\sin^2 \vartheta_{12}$  
for the MSW four--neutrino oscillations. 
The light (dark) regions are allowed at 90\% CL (99\% CL).}  
\label{fig:msw_rzs} 
\end{figure}

\begin{figure} 
\begin{center} 
\mbox{\epsfig{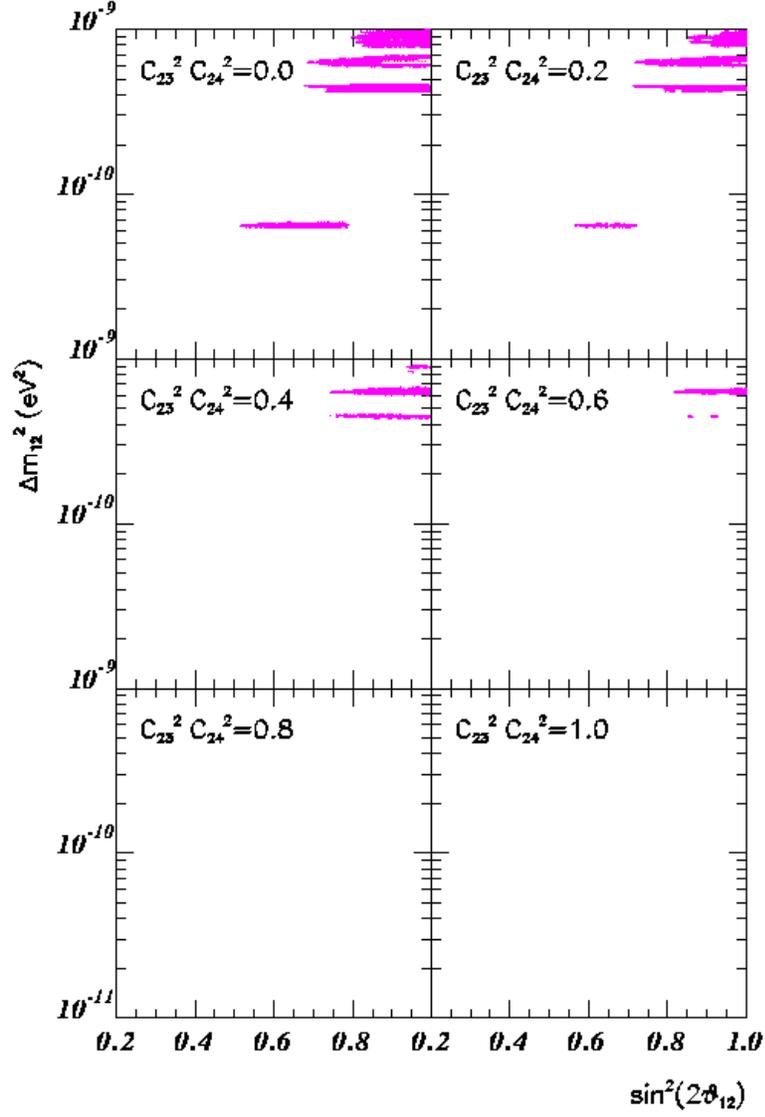}} 
\end{center} 
\caption{Results of the global analysis for the allowed regions in   
$\Delta{m}^2_{21}$ and $\sin^2(2\vartheta_{12})$  
for the vacuum four--neutrino oscillations. 
The light (dark) regions are allowed at 90\% CL (99\% CL).}  
\label{fig:vac_rzs} 
\end{figure}

We first present the results of the allowed regions in the three--parameter 
space for the different combination of observables. 
In building these regions, for a  
given set of observables,  we compute for any point in the parameter space  
of four--neutrino oscillations 
the expected values of the observables and with those and the corresponding 
uncertainties we construct the function  
$\chi^2(\Delta m_{12}^2,\vartheta_{12},c_{23}^2c_{24}^2)_{obs}$.  
We find its minimum in the full three-dimensional space considering as  
a unique framework both MSW and vacuum oscillations. The allowed regions  
for a given CL are then defined as the set of points satisfying  
the condition:  
\begin{equation} 
\chi^2(\Delta m_{12}^2,\vartheta_{12},c_{23}^2c_{24}^2)_{obs} 
-\chi^2_{min,obs}\leq \Delta\chi^2 \mbox{(CL, 3~dof)} 
\label{deltachi2} 
\end{equation}  
where, for instance, $\Delta\chi^2($CL, 3~dof)=6.25, 7.83, and 11.36 for 
CL=90, 95, and 99 \% respectively. In  
Figs.~\ref{fig:msw_r}--\ref{fig:vac_rzs} we plot the  
sections of such volume in the plane 
($\Delta{m}^2_{21},\sin^2(2\vartheta_{12})$) or  
($\Delta{m}^2_{21},\sin^2(\vartheta_{12})$) for different values of 
$c_{23}^2c_{24}^2$.  
 
\begin{figure} 
\begin{center} 
\mbox{\epsfig{file=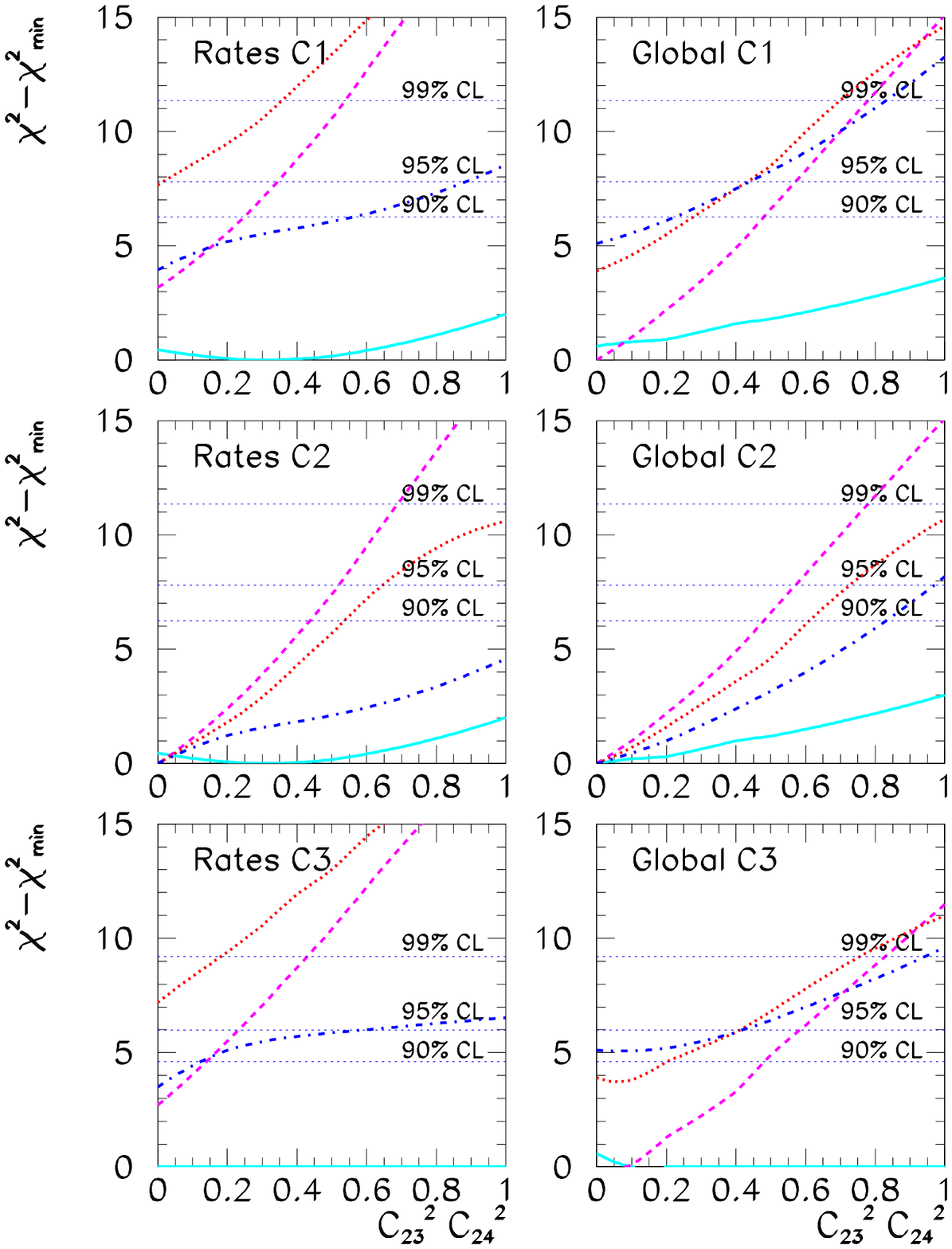,height=0.7\textheight}} 
\end{center} 
\caption{$\Delta \chi^2$ as a function of the mixing parameter 
$c_{23}^2 c_{24}^2$ for the different solutions SMA (full line), 
LMA (dashed), LOW(dotted) and vacuum (dot--dashed) from 
the analysis of the rates only right panels and from   
the analysis of the full data set (right panels).  
Each row represents the value for the different statistical  
criteria C1, C2 and C3 as defined in the text. The dotted 
horizontal lines correspond to the 90\%, 95 \%, 99\% CL  
limits for each criteria.} 
\label{chi2} 
\end{figure}

Figures~\ref{fig:msw_r} and ~\ref{fig:vac_r} show the results of  
the fit to the observed total rates only. We find that both at 90 and 99\% CL, 
the three--dimensional allowed volume is composed of three separated  
three-dimensional regions in the  MSW sector of the parameter space  
(Fig.~\ref{fig:msw_r}), 
which we denote as  SMA, LMA and LOW solutions following the  
usual  two--neutrino oscillation picture    
and a ``tower'' of regions in the vacuum oscillations sector  
(Fig.~\ref{fig:vac_r}). 
The values of the minimum of the $\chi^2$ 
in the different regions are given in Table~\ref{global}. The 
global minimum used in the construction of the volumes  
lies in the SMA region and for a non--vanishing 
value of $c_{23}^2c_{24}^2=0.3$, 
although, as can be seen in  
the first panel in Fig.~\ref{chi2}, this is of very little statistical 
significance as $\Delta\chi^2$ for the SMA solution is very mildly 
dependent on  $c_{23}^2c_{24}^2$   
($\Delta\chi^2 \lesssim 0.5$ for $c_{23}^2c_{24}^2\lesssim 0.5$). 
 
As seen in  Fig.~\ref{fig:msw_r}, 
the SMA region is always a valid solution  
for any value of $c_{23}^2c_{24}^2$. This is expected as  
in the two--neutrino oscillation picture this solution holds both  
for pure active--active and pure active--sterile oscillations\footnote{ 
Notice, however, that the statistical analysis is different: 
in the two--neutrino picture the pure active--active and active--sterile 
cases are analyzed separately, 
whereas in the four--neutrino picture they are taken into account 
simultaneously in a consistent scheme that allows to calculate 
the allowed regions 
with the prescription given in Eq.~(\ref{deltachi2}). 
We think that the agreement between the results of the analyses 
with two and four neutrinos indicate that the 
physical conclusions are quite robust. 
}. 
On the other hand, both the LMA and LOW solutions disappear for 
a value of the mixing $c_{23}^2c_{24}^2\gtrsim 0.5 (0.3)$.  
Unlike active neutrinos which lead to events in the 
water Cerenkov detectors by interacting via neutral current with the 
electrons, sterile neutrinos do not contribute to the Kamiokande and  
Super--Kamiokande event rates.  Therefore a larger survival probability  
for $^8B$ neutrinos is needed to accommodate the measured rate. As a  
consequence a larger contribution from $^8B$ neutrinos to the Chlorine  
and Gallium experiments is expected, so that the small measured rate in  
Chlorine can only be accommodated if no $^7Be$ neutrinos are present in the 
flux. This is only possible in the SMA solution region, since in the 
LMA and LOW regions the suppression of $^7Be$ neutrinos is not enough. 
 
In Table \ref{limits} we give the maximum values of  
$c_{23}^2c_{24}^2$  for which the different solutions are allowed 
at the 90 and 99\% CL according to different statistical criteria 
which we discuss below. In Fig.~\ref{fig:vac_r} we plot the corresponding 
sections in the vacuum oscillation sector. As seen in the figure,  
as $c_{23}^2c_{24}^2$  grows,   
the vacuum oscillation solution becomes more restricted in the  
allowed values of mass splittings till becoming a    
narrow band at $\Delta{m}^2_{21}\sim 10^{-10}$ eV$^2$ for the pure  
sterile case.  
 
Figure ~\ref{fig:msw_rz} shows the regions allowed by the fit of 
both total rates and the Super--Kamiokande zenith angular distribution 
in the MSW sector of the parameter space. In the  
vacuum oscillation section no day--night variation 
is expected. Also plotted is the excluded region at 99 \% CL from the 
zenith angular measurement. This exclusion volume is built as the 
corresponding three--degree--of--freedom region for the $\chi^2$ of the 
zenith angular data with respect to the minimum value $\chi^2_{min,zen}=0.8$ 
which occurs at $\Delta{m}^2_{21}=2.7\times 10^{-6}$ eV$^2$,  
$\sin^2(\vartheta_{12})=0.85$, and $c_{23}^2c_{24}^2=0.0$. We remark that  
this minimum is placed in the second octant and this was not included in  
past analysis of two-flavor MSW solutions although it leads to little 
effect in the final results of the allowed regions.  
As seen in the figure and also in  
Table \ref{limits}, the main effect of the inclusion of the day--night  
variation data is to cut down the lower part of the LMA region and to  
push towards slightly higher values the maximum  $c_{23}^2c_{24}^2$  
for which the LMA and the LOW solutions are still valid. 
 
In Figs.~\ref{fig:msw_rs} and~\ref{fig:vac_rs} we plot the regions allowed 
by the fit of both total rates and the Super--Kamiokande energy spectrum. 
Also plotted is the excluded region at 99 \% CL from the 
spectrum data which is obtained as the 
corresponding three--degree--of--freedom region for the $\chi^2$ of the 
spectrum data with respect to the minimum value $\chi^2_{min,spec}=15.1$ 
which occurs in the vacuum solution sector at  
$\Delta{m}^2_{21}=6.3\times 10^{-10}$ eV$^2$ and    
$\sin^2(2\vartheta_{12})=1$, and it is almost independent of  
$c_{23}^2c_{24}^2$.  As seen in the figure and also in  
Table \ref{limits}, the main effect of the inclusion of the spectrum data 
in the MSW regions is also to push towards slightly higher  
values the maximum  $c_{23}^2c_{24}^2$ for which the LMA and the LOW  
solutions are still valid. 
Figure \ref{fig:msw_rs} shows that the LMA 
region at 99 \% CL extends to high values of 
$\Delta{m}^2_{21}$, 
even above $10^{-3} \, \text{eV}^2$ 
for $c_{23}^2c_{24}^2 \lesssim 0.1$. 
Since the atmospheric mass squared difference 
$\Delta{m}^2_{\mathrm{atm}}$ 
lies between $10^{-3}$ and $10^{-2} \, \text{eV}^2$ 
(see \cite{SK-lp99}), 
one may wonder if the solar and atmospheric mass squared differences 
may coincide 
and three massive neutrinos may be enough for the explanation of 
solar, atmospheric and LSND data. 
The answer to this question is negative, 
because 
in the high--$\Delta{m}^2_{21}$ 
part of the 99 \% CL LMA region the mixing angle $\theta_{21}$ is large, 
$ 0.3 \lesssim \sin^2(\theta_{21}) \lesssim 0.7 $, 
and in this case disappearance of $\bar\nu_e$'s 
should be observed in long-baseline reactor experiments, 
contrary to results of the CHOOZ \cite{chooz} experiment. 
In other words, 
the results of the CHOOZ experiment, 
that have not been taken into account in the present analysis, 
forbid the part of the 99 \% CL LMA 
region that extends above 
$\Delta{m}^2_{21} \simeq 10^{-3} \, \text{eV}^2$.
For this reason we cut the plots at this value. 
For the vacuum sector, once the  
spectrum data is included the higher $\Delta{m}^2_{21}$ are favored but  
we find no region at the 90\% CL for any value of  
$c_{23}^2c_{24}^2\gtrsim 0.2$  
and at the 99\% CL the region totally disappears for  
$c_{23}^2c_{24}^2\gtrsim 0.8$.  
  
Figures~\ref{fig:msw_rzs} and~\ref{fig:vac_rzs} show the results  
from the global fit of the full set of data.  
The values of the minimum $\chi^2$ 
in the different regions are given in Table~\ref{global}. The 
global minimum used in the construction of the volumes  
lies in the LMA region and for vanishing $c_{23}^2 c_{24}^2$ 
corresponding to pure $\nu_e$--active neutrino oscillations. 
 
In Table~\ref{limits} we give the maximum values of $c_{23}^2c_{24}^2$   
for which the different solutions are allowed 
at the 90 and 99\% CL according to different statistical criteria. 
The use of each criteria depends on the physics scenario to which the  
result of our analysis is to be applied.  
\begin{itemize} 
\item Criterion 1 (C1): The maximum allowed value of the mixing  
$c_{23}^2c_{24}^2$  at a given CL for a given solution is defined 
as the value for which the corresponding region of the allowed  
three--dimensional volume defined as a 3-d.o.f shift with respect to the  
{\sl global minimum in the full parameter space}, disappears. In the first  
row in  Fig. \ref{chi2} we plot the values of $\Delta \chi^2$ defined in  
this way for the different 
solutions as a function of $c_{23}^2c_{24}^2$ for the fit of the total 
rates only (left panels) and for the global analysis (right panels).  
This criterion is the one used in building the regions in  
Figs.~\ref{fig:msw_r}--\ref{fig:vac_rzs}.  
It is applicable to models where no  
region of the parameter space MSW-SMA, MSW-LMA, MSW-LOW or vacuum  is 
favored.  
\item Criterion 2 (C2): The maximum allowed value of the mixing  
$c_{23}^2c_{24}^2$  at a given CL for a given solution is defined 
as the value for which the corresponding allowed three dimensional region  
defined as a 3-d.o.f shift with respect to the  
{\sl local minimum in the corresponding region}, disappears. In the second  
row in  Fig. \ref{chi2} we plot the values of $\Delta \chi^2$ defined in  
this way for the different 
solutions as a function of $c_{23}^2c_{24}^2$ for the fit to the total 
rates only (left panels) and for the global analysis (right panels).  
This criterion holds for models where only a certain   
solution, MSW-SMA, MSW-LMA, MSW-LOW or vacuum is possible and 
it yields less restrictive limits.  
\item Criterion 3 (C3): The maximum allowed value of the mixing  
$c_{23}^2c_{24}^2$ at a given CL is obtained
calculating the two-dimensional allowed regions
in the
$\sin^2\theta_{12}$--$\Delta m^2_{12}$
plane
for each fixed value of
$c_{23}^2c_{24}^2$.
These allowed regions are defined
through the 2-d.o.f shift with respect to the 
{\sl global minimum} in the plane $\sin^2\theta_{12}$--$\Delta m^2_{12}$
(this is the analogous of Criterion 1 for two parameters).
For each solution, the maximum allowed value of
$c_{23}^2c_{24}^2$
is that for which
the corresponding two dimensional region
in the $\sin^2\theta_{12}$--$\Delta m^2_{12}$
disappears.
This criterion is the equivalent to the usual two-neutrino analysis but 
with $\nu_e$ oscillating into a state which is a given superposition of 
active and sterile neutrino.

In the third row in Fig. \ref{chi2} we plot the difference
$\Delta \chi^2$
between the local minimum of $\chi^2$ for each solution and
the global minimum in the plane $\sin^2\theta_{12}$--$\Delta m^2_{12}$
as a function of $c_{23}^2c_{24}^2$.
Notice that for the analysis of rates only the minimum in the plane
$\sin^2\theta_{12}$ and $\Delta m^2_{12}$ occurs always in the MSW-SMA
region for any value of $c_{23}^2c_{24}^2$. Thefore the  
curve for the MSW-SMA solution corresponds to the horizontal 
$\Delta\chi^2=0$ line 
and it is not shown.
For the global analysis, when $c_{23}^2c_{24}^2<0.1$
the minimum in the plane occurs for the MSW-LMA (dashed line) 
solution while for  $c_{23}^2c_{24}^2>0.1$ it moves to the MSW-SMA (full line).
For this reason the curve for the MSW-SMA (MSW-LMA) solution is only seen 
for $c_{23}^2c_{24}^2<0.1$ ($>0.1$) while for $c_{23}^2c_{24}^2>0.1$ ($<0.1$)  
it coincides with the $\Delta\chi^2=0$ line.
In general this criterion is applicable for
models where the additional mixing $c_{23}^2c_{24}^2$ is fixed a priori 
to some value, so that the model in fact contains only two free parameters
(if $c_{23}^2c_{24}^2$ is larger than the maximum allowed for a given solution,
it means that that solution is not allowed in the specific model).

\end{itemize} 
 
\section{Summary and Discussion} 
\label{conclu} 
At present, 
the Standard Model assumption of massless neutrinos is  
under question due to the important results of underground experiments. 
Altogether they provide solid evidence for the existence of anomalies 
in the solar and atmospheric neutrino fluxes which could be accounted for 
in terms of neutrino oscillations $\nu_e \to \nu_x$ and $\nu_\mu \to \nu_x$, 
respectivelyly. 
Together with these results there is also the indication of 
neutrino oscillations in the $\bar\nu_\mu \rightarrow \bar\nu_e$ channel  
obtained in the LSND experiment.  
All these experimental results can be accommodated 
in a single neutrino oscillation framework only if there are at least 
three different scales of neutrino mass-squared differences. 
The simplest way to open the possibility of incorporating the LSND 
scale to the solar and atmospheric neutrino scales is to invoke a sterile  
neutrino, i.e. one whose interaction with 
Standard Model particles is much weaker 
than the SM weak interaction, 
so that it does not affect the invisible Z decay  
width, precisely measured at LEP. The sterile neutrino must also be light 
enough in order to participate in the oscillations involving the three 
active neutrinos. After imposing the present constrains from the negative  
searches at accelerator and reactor neutrino oscillation experiments 
one is left with two possible mass patterns which can be included in a 
single four--neutrino framework as described in  
Sec.~\ref{formalism}.  
 
In this paper we have performed  
an analysis of the neutrino oscillation  
solutions to the solar neutrino problem in the framework of  
four--neutrino mixing. We consider both transitions via the  
Mikheyev--Smirnov--Wolfenstein (MSW) mechanism as well as oscillations 
in  vacuum.  In what solar neutrinos is concerned 
our formalism contains one additional parameter 
as compared to the pure two--neutrino case:  
$\cos^2(\vartheta_{23}) \cos^2(\vartheta_{24})$, where  
$\vartheta_{23}$ and $\vartheta_{24}$  
give the projections of the sterile neutrino into each of the two heavier 
states responsible for explanation to the atmospheric neutrino anomaly.      
In this way, the formalism  permits transitions into active or sterile  
neutrino controlled by the additional parameter  
and contains as limiting cases the pure  $\nu_e$--active and  
$\nu_e$--sterile neutrino oscillations.  
 
We have studied the evolution of the different solutions to the 
solar neutrino problem in this three parameter space when the 
different set of observables are included.  
In Figs.~\ref{fig:msw_r}--\ref{fig:vac_rzs} we plot the  
sections of such volume in the plane 
($\Delta{m}^2_{21},\sin^2(2\vartheta_{12})$) or  
($\Delta{m}^2_{21},\sin^2(\vartheta_{12})$) for different values of 
$c_{23}^2c_{24}^2$. 
As a particularity, we also show that for MSW transitions there 
are solutions at 99 \% CL at $\vartheta_{12}$ mixing angles greater than  
$\frac{\pi}{4}$ and that the best fit point for the zenith angle distribution  
is in the second octant. 
 
Our results show that the SMA region is always a valid solution  
for any value of $c_{23}^2c_{24}^2$. This is expected as  
in the two--neutrino oscillation picture this solution holds both  
for pure active--active and pure active--sterile oscillations. 
On the other hand, the LMA, LOW and vacuum solutions become worse as  
the additional mixing $c_{23}^2c_{24}^2$ grows and they get to  
disappear for large values of the mixing.   
The main quantitative results of our analysis are summarized in  
Table~\ref{limits} and Fig.~\ref{chi2} where we give the maximum values of  
$c_{23}^2c_{24}^2$  for which the different solutions are allowed 
at the 90 and 99\% CL according to different statistical criteria 
which depend on  the physics scenario to which the  
result of our analysis is to be applied.  

\acknowledgments 
M.~C. G.-G. is thankful to the CERN theory division for their kind  
hospitality during her visit.  This work was supported 
by Spanish DGICYT under grant PB95-1077, by the European Union TMR 
network ERBFMRXCT960090. 
 
\newpage 
\appendix 
 
\section{Derivation of the relation between 
{\bf $P^{\text{E\lowercase{arth}}}_{\lowercase{\nu_2\to\nu_e}}$} 
and 
{\bf $P^{\text{E\lowercase{arth}}}_{\lowercase{\nu_2\to\nu_s}}$}} 
\label{append} 
 
In this appendix we derive the relation (\ref{relation}) 
between 
$P^{\text{Earth}}_{\nu_2\to\nu_e}$ 
and 
$P^{\text{Earth}}_{\nu_2\to\nu_s}$. 
In order to see the reason of this relation, 
it is useful to consider the most general $4\times4$ mixing matrix 
(without CP-violating phases) 
given in Eq.~(\ref{U4}), 
that can be written as 
\begin{equation} 
U = U' U_{12} 
\,, 
\label{Up1} 
\end{equation} 
with 
\begin{equation} 
U' 
= 
U_{34} \, U_{24} \, U_{23} \, U_{14} \, U_{13} 
\,. 
\label{Up} 
\end{equation} 

Let us define the neutrino states 
\begin{equation} 
|\nu'_r\rangle 
= 
\sum_{\alpha=e,s,\mu,\tau} U'_{{\alpha}r} |\nu_\alpha\rangle 
\qquad 
(r=1,2,3,4) 
\,. 
\label{nup} 
\end{equation} 

The amplitudes of 
$\nu_k\to\nu_\alpha$ 
transitions in the Earth 
for $k=1,2$ are given by 
\begin{equation} 
A^{\text{Earth}}_{\nu_k\to\nu_\alpha} 
= 
\langle \nu_\alpha | {\cal S} | \nu_k \rangle 
= 
\sum_{r=1}^4 
\langle \nu_\alpha | \nu'_r \rangle 
\langle \nu'_r | {\cal S} | \nu_k \rangle 
= 
\sum_{r=1}^4 
U'_{{\alpha}r} \, S'_{rk} 
\,, 
\label{02} 
\end{equation} 
where the unitary operator ${\cal S}$ 
describes the evolution inside the Earth 
and 
$ S'_{rk} \equiv \langle \nu'_r | {\cal S} | \nu_k \rangle $. 
 
It has been shown in Ref.~\cite{DGKK-99} 
that the matter effects inside the Earth can generate 
only transitions between 
$\nu_1,\nu_2$ 
and 
$\nu'_1,\nu'_2$. 
Then, 
we have 
\begin{eqnarray} 
& 
A^{\text{Earth}}_{\nu_k\to\nu_\alpha} 
= 
U'_{\alpha1} \, S'_{1k} + U'_{\alpha2} \, S'_{2k} 
& 
\qquad 
(k = 1, 2) 
\,, 
\label{03} 
\\ 
& 
A^{\text{Earth}}_{\nu_k\to\nu_\alpha} 
= 
U'_{{\alpha}k} 
= 
U_{{\alpha}k} 
& 
\qquad 
(k = 3, 4) 
\,, 
\label{04} 
\end{eqnarray} 
and 
the transition probabilities are given by 
\begin{eqnarray} 
& 
P^{\text{Earth}}_{\nu_k\to\nu_\alpha} 
= 
U^{\prime2}_{\alpha1} \, |S'_{1k}|^2 
+ 
U^{\prime2}_{\alpha2} \, |S'_{2k}|^2 
+ 2 U'_{\alpha1} \, U'_{\alpha2} \, \text{Re}[ S'_{1k} S^{\prime*}_{2k} ] 
& 
\qquad 
(k = 1, 2) 
\,, 
\label{05} 
\\ 
& 
P^{\text{Earth}}_{\nu_k\to\nu_\alpha} 
= 
U_{{\alpha}k}^2 
& 
\qquad 
(k = 3, 4) 
\,. 
\label{06} 
\end{eqnarray} 
 
Since the evolution operator ${\cal S}$ is unitary, 
the matrix $S'$ is unitary 
and 
we have the relations 
\begin{equation} 
|S'_{12}|^2 = |S'_{21}|^2 \equiv P'_{12} 
\,, 
\quad 
|S'_{11}|^2 = |S'_{22}|^2 = 1 - P'_{12} 
\,, 
\quad 
S'_{11} \, S^{\prime*}_{21} + S'_{12} \, S^{\prime*}_{22} = 0 
\,, 
\label{0203} 
\end{equation} 
where $P'_{12}$ is the probability of 
$\nu_2\leftrightarrows\nu'_1$ 
transitions, 
that is equal to 
the probability of 
$\nu_1\leftrightarrows\nu'_2$ 
transitions. 
It is easy to check that the unitarity constraints 
(\ref{0203}) 
are equivalent to the probability conservation relations 
\begin{equation} 
\sum_{k=1}^4 P^{\text{Earth}}_{\nu_k\to\nu_\alpha} = 1 
\,, 
\qquad 
\sum_{\alpha=e,s,\mu,\tau} P^{\text{Earth}}_{\nu_k\to\nu_\alpha} = 1 
\,. 
\label{08} 
\end{equation} 
 
Let us notice that 
by construction the matrix $U'$ is such that 
\begin{equation} 
U'_{e2} = 0 
\,, 
\label{091} 
\end{equation} 
and 
the probability of 
$\nu_k\to\nu_e$ 
transitions inside the Earth depend only on 
$U^{\prime2}_{e1} = \cos\vartheta_{13}^2 \cos\vartheta_{14}^2 $ 
and 
$P'_{12}$: 
\begin{equation} 
P^{\text{Earth}}_{\nu_1\to\nu_e} 
= 
U^{\prime2}_{e1} \left( 1 - P'_{12} \right) 
\, 
\qquad 
P^{\text{Earth}}_{\nu_2\to\nu_e} 
= 
U^{\prime2}_{e1} \, P'_{12} 
\,. 
\label{093} 
\end{equation} 

On the other hand, 
in general, 
for a given mixing matrix $U$, 
the transition probabilities 
$P^{\text{Earth}}_{\nu_k\to\nu_\alpha}$ 
with 
$\alpha=s,\mu,\tau$ 
depend on two independent quantities, 
$P'_{12}$ and $\text{Arg}[S'_{11}S^{\prime*}_{21}]$. 
Hence, 
in general 
the probabilities 
$P^{\text{Earth}}_{\nu_k\to\nu_e}$ 
and 
$P^{\text{Earth}}_{\nu_k\to\nu_s}$ 
must be calculated independently. 
However, 
if  
\begin{equation} 
U'_{\alpha1} \, U'_{\alpha2} = 0 
\qquad 
(\alpha=s,\mu,\tau) 
\label{094} 
\,, 
\end{equation} 
also $P^{\text{Earth}}_{\nu_k\to\nu_\alpha}$ 
depends only on the elements of the mixing matrix and 
$P'_{12}$: 
\begin{equation} 
P^{\text{Earth}}_{\nu_1\to\nu_\alpha} 
= 
U^{\prime2}_{\alpha1} \left( 1 - P'_{12} \right) 
+ 
U^{\prime2}_{\alpha2} \, P'_{12} 
\,, 
\qquad 
P^{\text{Earth}}_{\nu_2\to\nu_\alpha} 
= 
U^{\prime2}_{\alpha1} \, P'_{12} 
+ 
U^{\prime2}_{\alpha2} \left( 1 - P'_{12} \right) 
\,, 
\label{096} 
\end{equation} 
for $\alpha=s,\mu,\tau$. 
 
In the approximation 
$\vartheta_{13}=\vartheta_{14}=0$ 
that we use in the analysis of solar neutrino data, 
we have 
\begin{equation} 
U'_{e1} = 1 
\, 
\qquad 
\qquad 
U'_{\alpha1} = 0 
\qquad 
(\alpha=s,\mu,\tau) 
\,. 
\label{12} 
\end{equation} 

Therefore, 
the condition (\ref{094}) is satisfied and we obtain 
\begin{eqnarray} 
&& 
P^{\text{Earth}}_{\nu_1\to\nu_e} 
= 
1 - P'_{12} 
\, 
\qquad 
P^{\text{Earth}}_{\nu_2\to\nu_e} 
= 
P'_{12} 
\label{121a} 
\\ 
&& 
P^{\text{Earth}}_{\nu_1\to\nu_\alpha} 
= 
U^{\prime2}_{\alpha2} \, P'_{12} 
\,, 
\qquad 
P^{\text{Earth}}_{\nu_2\to\nu_\alpha} 
= 
U^{\prime2}_{\alpha2} \left( 1 - P'_{12} \right) 
\,. 
\label{121b} 
\end{eqnarray} 

Eliminating $P'_{12}$ 
from the relations (\ref{121a}) and (\ref{121b}), 
we obtain 
\begin{equation} 
P^{\text{Earth}}_{\nu_k\to\nu_\alpha} 
= 
U^{\prime2}_{\alpha2} 
\left( 1 - P^{\text{Earth}}_{\nu_k\to\nu_e} \right) 
\qquad 
( k=1,2 ; \, \alpha=s,\mu,\tau ) 
\,. 
\label{26} 
\end{equation} 

In particular, 
for $k=2$ and $\alpha=s$, 
we obtain the useful relation (\ref{relation}) between 
$P^{\text{Earth}}_{\nu_2\to\nu_s}$ 
and 
$P^{\text{Earth}}_{\nu_2\to\nu_e}$. 
 
In general, 
considering the possibility of small but non-zero 
$\vartheta_{13}$ and/or $\vartheta_{14}$, 
if the mixing angles are such that $U'_{\alpha1}=0$ 
for $\alpha=s,\mu,\tau$, 
we have the relation 
\begin{equation} 
P^{\text{Earth}}_{\nu_k\to\nu_\alpha} 
= 
U^{\prime2}_{\alpha2} 
\left( 
1 - 
\frac{ P^{\text{Earth}}_{\nu_k\to\nu_e} }{ U^{\prime2}_{e1} } 
\right) 
\qquad 
(\alpha=s,\mu,\tau) 
\,, 
\label{32} 
\end{equation} 
whereas if $U'_{\alpha2}=0$ we obtain the relation 
\begin{equation} 
P^{\text{Earth}}_{\nu_k\to\nu_\alpha} 
= 
\frac{ U^{\prime2}_{\alpha1} }{ U^{\prime2}_{e1} } 
P^{\text{Earth}}_{\nu_k\to\nu_e} 
\qquad 
(\alpha=s,\mu,\tau) 
\,. 
\label{33} 
\end{equation}

\newpage 
 
\begin{table} 
\begin{tabular}{|l|l|l|l|l|} 
Experiment & Rate & Ref. & Units& $ R^{\text BP98}_i $\\ 
\hline 
Homestake  & $2.56\pm 0.23 $ & \protect\cite{homestake} & SNU &  $7.8\pm 1.1 $   \\ 
GALLEX + SAGE  & $72.3\pm 5.6 $ & \protect\cite{gallex,sage} & SNU & $130\pm 7 $  \\ 
Kamiokande & $2.80\pm 0.38$ & \protect\cite{kamioka} &  
$10^{6}$~cm$^-2$~s$^{-1}$ & $5.2\pm 0.9 $ \\    
Super--Kamiokande & $2.45\pm 0.08$ & \protect\cite{sk99} &  
$10^{6}$~cm$^{-2}$~s$^{-1}$ & $5.2\pm 0.9 $ \\    
\end{tabular} 
\vglue .3cm 
\caption{Measured rates for the Chlorine, Gallium, Kamiokande and Super--Kamiokande  
experiments. } 
\label{rates} 
\end{table} 
\begin{table} 
\begin{tabular}{|c|c|c|c|c|c|} 
\hline 
& & rates & rates+zenith  & rates+spectrum & global \\ 
\hline 
& $\Delta m^2$                 &$5.6\times 10^{-6}$   & 
$6.2\times 10^{-6}$ & 
$5.2\times 10^{-6}$ & 
$5.2\times 10^{-6}$    \\ 
SMA & $\sin^2(2\vartheta)$        &$6.0\times 10^{-3}$   & 
$5.4\times 10^{-3}$ & 
$4.7\times 10^{-3}$ & 
$4.7\times 10^{-3}$    \\ 
    & $c_{23}^2 c_{24}^2$      &0.3   &0.4   & 0.0  & 0.0 \\ 
& $\chi^2_{min}$ (Prob \%)   & 0.55 (45) &6.2 (40)  & 23.9 (16) & 29.7 (16)\\ 
\hline   
& $\Delta m^2$                 &$1.4\times 10^{-5}$   & 
$4.2\times 10^{-5}$   & 
$1.4\times 10^{-5}$   & 
$3.9\times 10^{-5}$    \\ 
LMA & $\sin^2(2\vartheta)$        & 0.68   &0.81   &0.68   &0.78    \\ 
    & $c_{23}^2 c_{24}^2$      & 0.0   &0.0   & 0.0  & 0.0   \\ 
& $\chi^2_{min}$ (Prob \%)     & 3.80 (5)  &8.6 (20)   &23.1 (19)    
&29.1 (18)    \\ 
\hline   
& $\Delta m^2$                 &  
$1.3\times 10^{-7}$  & 
$1.1\times 10^{-7}$   & 
$1.0\times 10^{-7}$   & 
$1.0\times 10^{-7}$    \\ 
LOW & $\sin^2(2\vartheta)$        &0.93   &0.93   &0.93   &0.93    \\ 
    & $c_{23}^2 c_{24}^2$      &0.0   &0.0   &0.0   &0.0   \\ 
& $\chi^2_{min}$ (Prob \%)     &8.3 (0.4) &13.6 (3.4)    
&27.9 (6.4)   & 33.0 (8.1)   \\ 
\hline   
& $\Delta m^2$                 & 
$9.1\times 10^{-10}$   & 
$9.1\times 10^{-10}$   & 
$4.5\times 10^{-10}$   & 
$4.4\times 10^{-10}$    \\ 
Vacuum & $\sin^2(2\vartheta)$     &0.78   &0.78   &0.9   &0.9    \\ 
       & $c_{23}^2 c_{24}^2$   &0.0    &0.0    &0.0   &0.0    \\ 
       & $\chi^2_{min}$ (Prob \%)  
       & 4.5 (3.4)   &9.9 (13)  &28.8 (5.1)   &34.3 (6.1)    \\ 
\hline 
\end{tabular} 
\caption{Best fit points and the corresponding probabilities for the 
different solutions to the solar neutrino deficit  
and for different combinations of observables.} 
\label{global} 
\end{table} 
\begin{table} 
\begin{tabular}{|c|c|c|c|c|}\hline 
Observ. & Crit. & LMA 90 (99)  &  LOW 90 (99) & VAC 90 (99)  \\ 
\hline 
 &   C1 &  0.25 (0.54) &  --  (0.36) &   0.56 (1.0) \\ 
     Rates &   C2 &  0.44 (0.69) &  0.54 (1.00) &  1.0 (1.0) \\ 
     &   C3 &  0.14 (0.43) &  -- (0.18) &  0.12 (1.0) \\ 
\hline 
     &   C1 &  0.30 (0.61) &  -- (0.45) &   0.61 (1.0)\\ 
 Rates+SK Zenith &   C2 &  0.45 (0.74) &  0.59 (0.90) &  1.0 (1.0)\\ 
     &   C3 &  0.19 (0.49) & --  (0.23) &  0.15 (1.0) \\ 
\hline 
     &   C1 &  0.43 (0.67)&  0.18 (0.56) &  0.12 (0.77) \\ 
 Rates+SK Spect &   C2 &  0.43 (0.67) &  0.54 (1.0) & 0.83 (1.0) \\ 
     &   C3 &  0.45 (0.72) &  -- (0.57) &  -- (0.93) \\ 
\hline 
     &   C1 &  0.48 (0.78) &  0.28 (0.70) &  0.22 (0.83) \\ 
     Global fit &   C2 &  0.48 (0.78) &  0.61 (1.0) &  0.83 (1.0) \\ 
     &   C3 &  0.48 (0.83) &  0.20 (0.76) &  -- (0.95) \\ 
\hline 
\end{tabular} 
\caption{Maximum allowed value of $\cos^2_{23} \cos^2_{24}$ at 90 \% 
and 99 \% CL for the different solutions to the solar neutrino problem  
with the different statistical criteria. 
SMA is allowed at 90 \% for all the  
range of $\cos^2_{23} \cos^2_{24}$.} 
\label{limits} 
\end{table} 
%
%

\begin{thebibliography}{99} 
%
\bibitem{homestake0} 
R. Davis, Jr, D. S. Harmer, and K. C. Hoffman, Phys. Rev. Lett. {\bf 
20}, 1205 (1968) 
%
\bibitem{SSMold} 
J. N. Bahcall, N. A. Bahcall, and G. Shaviv, Phys. Rev. Lett. {\bf 
20}, 1209 (1968); J. N. Bahcall, R. Davis, Jr., Science {\bf 191}, 264 
(1976). 
%
\bibitem{Bahcall:1997qw} 
J.N.~Bahcall, M.H.~Pinsonneault, S.~Basu and J.~Christensen-Dalsgaard, 
Phys. Rev. Lett. {\bf 78}, 171 (1997) 
 
\bibitem{Bahcall:1995bt} 
J.N.~Bahcall and M.H.~Pinsonneault, Rev. Mod. Phys. {\bf 67}, 781 
(1995) 
 
\bibitem{homestake} 
B.~T.~Cleveland {\em et al.}, Ap.~J. {\bf 496}, 505 (1998).  
 
\bibitem{gallex} 
T.~Kirsten, Talk at the Sixth international workshop on topics in 
astroparticle and underground physics September, TAUP99, Paris, 
September 1999. 
 
\bibitem{sage}  
SAGE Collaboration, J. N. Abdurashitov et al., Phys. Rev. {\bf C60}, 
055801 (1999).
 
\bibitem{kamioka}  
Kamiokande Collaboration, Y. Fukuda et al., Phys. Rev. Lett. {\bf 77}, 
1683 (1996). 
 
\bibitem{sk1}  
Super--Kamiokande Collaboration, Y. Fukuda et al., 
Phys. Rev. Lett. {\bf 82}, 1810 (1999). 
Super--Kamiokande Collaboration, Y. Fukuda et al., 
Phys. Rev. Lett. {\bf 82}, 2430 (1999). 
 
\bibitem{sk99}  
Y. Suzuki, talk a the ``XIX International Symposium on 
Lepton and Photon Interactions at High Energies'', Stanford University,  
August 9-14, 1999;  
M. Nakahata, talk at the  
``6th International Workshop on Topics in Astroparticle and  
Underground Physics, TAUP99'', Paris, September 1999. 
 
\bibitem{ourseasonal} 
P. C. de Holanda, C. Pe\~na-Garay, M.\ C.\ Gonzalez-Garcia and J.\ W.\ 
F.\ Valle, Phys. Rev. {\bf D60}, 093010 (1999)
\bibitem{v99} 
Talks by M.~C.~Gonzalez-Garcia and A. Yu. Smirnov, Proceedings of {\sl 
International Workshop on Particles in Astrophysics and Cosmology: on 
Theory to Observation} Valencia, May 3-8, 1999, to be published by 
Nucl. Phys. Proc. Supplements, ed. V. Berezinsky, G. Raffelt and 
J. W.~F. Valle (http://flamenco.uv.es//v99.html) 
 
\bibitem{Glashow:1987jj} 
V.N.~Gribov and B.M.~Pontecorvo, Phys. Lett. {\bf 28B}, 493 (1969); 
V. Barger, K. Whisnant, R.J.N. Phillips, Phys. Rev. {\bf D24}, 538 
(1981); S.L.~Glashow and L.M.~Krauss, Phys. Lett. {\bf 190B}, 199 
(1987); V.~Barger, R.J.~Phillips and K.~Whisnant, 
Phys. Rev. Lett. {\bf 65}, 3084 (1990); S.L.~Glashow, P.J.~Kernan and 
L.M.~Krauss, Phys. Lett. {\bf B445}, 412 (1999); V. Berezinsky, 
G.~Fiorentini and M.~Lissia, hep-ph/9811352 and hep-ph/9904225. 
 
\bibitem{msw} 
S.P. Mikheyev and A.Yu. Smirnov, Sov. Jour. Nucl. Phys.  
42, 913 (1985); L.\ Wolfenstein, Phys.\ Rev.\ {\bf D17}, 2369 (1978). 
 
\bibitem{atmexp} NUSEX Collab., M. Aglietta {\sl et al.}, 
Europhys.  Lett.  {\bf 8}, 611 (1989); Fr\'ejus Collab., Ch. 
Berger {\sl et al.}, Phys.  Lett.  {\bf B227}, 489 (1989); IMB 
Collab., D. Casper {\sl et al.}, Phys. Rev. Lett.  {\bf 66}, 
2561 (1991); R. Becker-Szendy {\sl et al.}, Phys. Rev. {\bf D46}, 3720 
(1992); Kamiokande Collab., H. S. Hirata {\sl et al.}, 
Phys. Lett. {\bf B205}, 416 (1988) and Phys. Lett. {\bf B280}, 146 
(1992); Kamiokande Collab., Y. Fukuda {\sl et al.}, Phys. 
Lett. {\bf B335}, 237 (1994); Soudan Collab., W.  W.  M Allison 
{\sl et al.}, Phys.  Lett.  {\bf B391}, 491 (1997). 
 
\bibitem{atmSK} 
Super--Kamiokande Collaboration, Y. Fukuda et al., 
Phys. Rev. Lett. {\bf 81}, 1582 (1998).
Super--Kamiokande Collaboration, Y. Fukuda et al., 
Phys. Rev. Lett. {\bf 82}, 2644 (1999).
 
\bibitem{lsnd} C. Athanassopoulos, Phys.\ Rev.\ Lett.\ {\bf 75} 2650 (1995);  
Phys.\ Rev.\ Lett.\ {\bf 77} 3082 (1996);  
Phys.\ Rev.\ Lett.\ {\bf 81} 1774 (1998). 
 
\bibitem{four-models} 
J.T. Peltoniemi, D. Tommasini and J.W.F. Valle, Phys. Lett.{\bf B298}, 383 
(1993); E.J. Chun {\it et al.}, Phys. Lett.{\bf B357}, 608 (1995); 
S.C. Gibbons {\it et al.}, Phys. Lett. {\bf B430}, 296 (1998); B. 
Brahmachari and R.N. Mohapatra, Phys. Lett. {\bf B437}, 100 (1998); S. 
Mohanty, D.P. Roy and U. Sarkar, Phys. Lett. {\bf B445}, 185 (1998); J.T. 
Peltoniemi and J.W.F. Valle, Nucl. Phys. {\bf B406}, 409 (1993); Q.Y. Liu 
and A.Yu. Smirnov, Nucl. Phys. {\bf B524}, 505 (1998); 
D.O.~Caldwell, hep-ph/9804367 (to be publ. in Proc. Cosmo'97); 
D.O. Caldwell and 
R.N. Mohapatra, Phys. Rev. {\bf D48}, 3259 (1993); E. Ma and P. Roy, 
Phys. Rev. {\bf D52}, R4780 (1995); A.Yu. Smirnov and M. Tanimoto, Phys. 
Rev. {\bf D55}, 1665 (1997); N. Gaur {\it et al.}, Phys. Rev.  
{\bf D58}, 071301 (1998); E.J. Chun, C.W. Kim and U.W. Lee, Phys. Rev.  
{\bf D58}, 093003 (1998); K. Benakli and A.Yu. Smirnov, Phys. Rev. Lett. 
{\bf 79}, 4314 (1997); Y. Chikira, N. Haba and Y. Mimura, hep-ph/9808254; 
C. Liu and J. Song, Phys. Rev. {\bf D60}, 036002 (1999); 
W. Grimus, R. Pfeiffer and T. Schwetz, hep-ph/9905320. 
 
\bibitem{four-phenomenology} 
J.J. Gomez-Cadenas and M.C. Gonzalez-Garcia, Z. Phys. {\bf C71}, 443 
(1996); N. Okada and O. Yasuda, Int. J. Mod. Phys.  
{\bf A12}, 3669 (1997); S. Goswami, Phys. Rev.  
{\bf D55}, 2931 (1997); 
S.~M. Bilenky, C.~Giunti, and W.~Grimus, 
Phys. Rev. {\bf D57}, 1920 (1998); 
S.~M. Bilenky, C.~Giunti, and W.~Grimus, 
Phys. Rev. {\bf D58}, 033001 (1998); 
S.M. Bilenky, C. Giunti, W. Grimus 
and T. Schwetz, Astropart. Phys. {\bf 11}, 413 (1999); V. 
Barger, Y.B. Dai, K. Whisnant and B.L. Young, Phys. Rev. D {\bf 59}, 
113010 (1999); 
V. Barger, T.J. Weiler and K. Whisnant, Phys. 
Lett. B {\bf 427}, 97 (1998); 
C. Giunti, 
Phys. Rev. D {\bf 61}, 036002 (2000); 
C. Giunti, 
Phys. Lett. B {\bf 467}, 83 (1999); 
S.M. Bilenky {\it et al.}, 
Phys. Lett. B {\bf 465}, 193 (1999); 
C. Giunti, 
hep-ph/9912211; 
A. Ibarra and I. Navarro, 
hep-ph/9912282. 
 
\bibitem{BGG-AB} 
S.M. Bilenky, C. Giunti and W. Grimus, Eur. Phys. J. C {\bf 1}, 247 (1998), 
Proc. of {\it Neutrino '96}, Helsinki, June 1996, edited 
by K. Enqvist {\it  et al.}, p.~174, World Scientific, 1997, 
hep-ph/9609343; 
S.~M. Bilenky, C.~Giunti, W.~Grimus, and T.~Schwetz, 
Phys. Rev. {\bf D60}, 073007 (1999). 
 
\bibitem{Barger-variations-98} 
V.~Barger, S.~Pakvasa, T.~J. Weiler, and K.~Whisnant, 
Phys. Rev. {\bf D58}, 093016 (1998). 
 
\bibitem{DGKK-99} 
D. Dooling, C. Giunti, K. Kang and C.W. Kim, hep-ph/9908513. 
 
\bibitem{daynight} 
J. Bouchez {\it et. al.}, Z. Phys. {\bf C32}, 499 (1986);
S. P. Mikheyev and A. Yu. Smirnov, {\it '86 Massive Neutrinos in
Astrophysics and in Particle Physics}, proceedings of the Sixth
Moriond Workshop, edited by O. Fackler and J. Tr$\hat{a}$n Thanh
V$\hat{a}$n (Editions Fronti\`eres, Gif-sur-Yvette, 1986), pp. 355;
S.P. Mikheyev and A.Yu. Smirnov, Sov. Phys. Usp. 30 (1987) 759-790;
A. Dar {\it et. al.} Phys. Rev. {\bf D 35} (1987) 3607;
 E. D. Carlson,Phys. Rev. {\bf D34}, 1454 (1986) ; 
A.J. Baltz and J. Weneser,Phys. Rev. {\bf D50}, 5971 (1994);
 A. J. Baltz and J. Weneser,Phys. Rev. {\bf D51}, 3960 (1994); 
P. I. Krastev, hep-ph/9610339;
Q.Y. Liu, M. Maris and S.T. Petcov, Phys. Rev. {\bf D56}, 5991 (1997);
M. Maris and S.T. Petcov, Phys. Rev. {\bf D56}, 7444 (1997);
J.N. Bahcall and P.I. Krastev, Phys. Rev. {\bf C56}, 2839 (1997);
A. J. Baltz and J. Weneser, Phys. Rev. {\bf D35}, 528 (1987);
A. J. Baltz and J. Weneser, Phys. Rev. {\bf D37}, 3364 (1988);
E.~Lisi and D.~Montanino, Phys. Rev. {\bf D56}, 1792 (1997);
S.~T. Petcov, Phys. Lett. {\bf B434}, 321 (1998);
M.~Chizhov, M.~Maris, and S.~T. Petcov, (1998), hep-ph/9810501; 
M.V. Chizhov and S.T. Petcov, Phys. Rev. Lett. {\bf 83}, 1096 (1999);
A.S. Dighe, Q.Y. Liu and A.Yu. Smirnov, hep-ph/9903329; 
A.H. Guth, L. Randall and M. Serna, J. High Energy Phys. {\bf 8}, 018 (1999).

\bibitem{Akhmedov-parametric-99} 
E.~K. Akhmedov, Nucl. Phys. {\bf B538}, 25 (1999).

\bibitem{BGG-review-98} 
S.M. Bilenky, C. Giunti, and W. Grimus, 
Prog. Part. Nucl. Phys. {\bf 43}, 1 (1999). 

\bibitem{lisi3} G.L. Fogli, E. Lisi, D. Montanino and A. Palazzo,  
hep-ph/9912231. 
 
\bibitem{BP98}  
J.N. Bahcall, S. Basu and M. Pinsonneault, Phys. Lett. B433 (1998) 1. 
 
\bibitem{prod}  
http://www.sns.ias.edu/\~{}jnb/SNdata 
 
\bibitem{Faid}  
B.\ Fa\"{\i}d, G.\ L.\ Fogli, E.\ Lisi and D.\ Montanino, 
Phys.\ Rev.\ {\bf D55}, 1353 (1997). 
%
\bibitem{CrSe} 
J. N. Bahcall, M.\ Kamionkowsky, and A.\ Sirlin, Phys.\ Rev.\ {\bf 
D51}, 6146 (1995). 
 
\bibitem{fogli-lisi}  
G.\ L.\ Fogli, E.\ Lisi and D.\ Montanino, Phys.\ Rev.\ D {\bf 49}, 
3226 (1994). G.\ L.\ Fogli, E.\ Lisi, Astropart. Phys.{\bf 3}, 185 
(1995). 
 
\bibitem{oursolar} M.C. Gonzalez-Garcia, P.C. de Holanda, C. Pe\~na-Garay and 
J.W.F. Valle, hep-ph/9906469, to appear in Nucl. Phys. {\bf B}. 
 
 
\bibitem{bugey} B.\ Achkar {\sl et al.}, Nucl.\ Phys.\ {\bf B424},
 503  (1995). 
%
\bibitem{chooz} 
CHOOZ Collaboration, M. Apollonio {\sl et al.}.  
Phys.\ Lett.\ {\bf B420}, 397 (1998). 
 
 
\bibitem{Murayama} 
Andre de Gouvea, Alexander Friedland and Hitoshi Murayama, 
 hep-ph/9910286. 
 
\bibitem{SK-lp99} 
M. C. Gonzalez-Garcia, H. Nunokawa, O. L. G. Peres, T. Stanev and
J. W. F. Valle, Phys. Rev. {\bf D58}, 033004 (1998);
M.C.~Gonzalez-Garcia, H.~Nunokawa, O.L.~Peres and J.~W.~F.~Valle,
Nucl. Phys. {\bf B543}, 3 (1999); 
for a recent experimental review, see W. Anthony Mann, 
Plenary talk at the XIX Int. Symposium on Lepton and Photon 
Interactions at High Energies, Stanford, Aug. 1999, 
hep-ex/9912007. 
 
\end{thebibliography}
\end{document}